\documentclass[preprintnumbers,amsmath,amssymb,prd, notitlepage,nofootinbib,twocolumn, superscriptaddress]{revtex4}

\usepackage{graphicx}
\usepackage{dcolumn}
\usepackage{bm}
\usepackage{subfigure}
\usepackage{wasysym}
\usepackage{verbatim}
\usepackage{color}

\newcommand{\be}{\begin{equation}}
\newcommand{\ee}{\end{equation}}

\newcommand{\dd}{\mathrm{d}}
\newcommand{\st}{\sigma_\mathrm{T}}
\newcommand{\me}{m_\mathrm{e}}
\newcommand{\nume}{n_\mathrm{e}}
\newcommand{\rmn}{\mathrm}
\newcommand{\N}{\mathcal{N}}

\newcommand{\Mnu}{\Sigma m_{\nu}}
\newcommand{\tht}{\bm{\theta}}

\begin{document}

\title{Fundamental Physics from Future Weak-Lensing Calibrated Sunyaev-Zel'dovich Galaxy Cluster Counts}

\author{Mathew S. Madhavacheril}
\affiliation{Department of Astrophysical Sciences, Princeton University, Princeton, NJ 08544, USA}
\email{mathewm@astro.princeton.edu}

\author{Nicholas Battaglia}
\affiliation{Department of Astrophysical Sciences, Princeton University, Princeton, NJ 08544, USA}
\affiliation{Department of Astronomy, Cornell University, Ithaca, NY 14853, USA}
\affiliation{Center for Computational Astrophysics, Flatiron Institute, 162
Fifth Avenue, New York, NY 10010, USA}

\author{Hironao Miyatake}
\affiliation{Jet Propulsion Laboratory, California Institute of Technology, Pasadena, CA 91109, USA}
\affiliation{Kavli Institute for the Physics and Mathematics of the Universe (Kavli IPMU, WPI), UTIAS, Tokyo Institutes for Advanced Study, The University of Tokyo, Chiba 277-8583, Japan}

\date{\today}

\begin{abstract}

Future high-resolution measurements of the cosmic microwave background (CMB) will produce catalogs of tens of thousands of galaxy clusters through the thermal Sunyaev-Zel'dovich (tSZ) effect. We forecast how well different configurations of a CMB Stage-4 experiment can constrain cosmological parameters, in particular the amplitude of structure as a function of redshift $\sigma_8(z)$, the sum of neutrino masses $\Mnu$, and the dark energy equation of state $w(z)$. A key element of this effort is calibrating the tSZ scaling relation by measuring the lensing signal around clusters. We examine how the mass calibration from future optical surveys like the Large Synoptic Survey (LSST) compares with a purely internal calibration using lensing of the CMB itself. We find that, due to its high-redshift leverage, internal calibration gives constraints on cosmological parameters comparable to the optical calibration, and can be used as a cross-check of systematics in the optical measurement. We also show that in contrast to the constraints using the CMB lensing power spectrum, lensing-calibrated tSZ cluster counts can detect a minimal $\Mnu$ at the 3-5$\sigma$ level even when the dark energy equation of state is freed up. 

\end{abstract}

\maketitle

\section{Introduction}

The abundance of galaxy clusters is a sensitive probe of the amplitude of density fluctuations that scales strongly with the normalization of the matter power spectrum, $\sigma_8$, and the matter density, $\Omega_m$ \citep[e.g.,][]{Voit2005,AEM2011}. Measuring cluster abundance as a function of redshift allows one to probe physics that affects the growth of structure, for example the effect of massive neutrinos and the dark energy equation of state. Recent constraints from measurements of cluster abundances have however been limited by systematic effects \citep[e.g.,][]{2009ApJ...692.1060V,Vand2010,Sehgal2011,Benson_2013,Hass2013,Planckcounts,Mantz2014,PlnkSZCos2015,Mantz2015,dehaan2016}, the dominant systematic uncertainty being the calibrations of observable-to-mass relations. Therefore, accurate and precise calibrations of the observable-to-mass relation is essential for any future cluster cosmological constraint.

Galaxy clusters are observationally identified across the electromagnetic spectrum, from microwaves to X-ray energies. The measurements of secondary temperature anisotropies in the CMB that arise from the tSZ effect \citep{SZ1970} are emerging as a powerful tool to find and count clusters. To compare with observational abundances the theoretical abundance predictions are typically forward modeled from cosmological parameters through a mass function \citep[e.g.,][]{PS1974,Tink2008} and an observable--mass relation. For example, the recent Planck \citep{PlnkSZCos2015} and South Pole Telescope \citep[SPT,][]{dehaan2016} cosmological constraints from tSZ cluster counts included weak-lensing and X-ray masses to calibrate their observable--mass relation, and Atacama Cosmology Telescope (ACT) used velocity dispersions \citep{Hass2013}. In all tSZ cluster count analyses an independent method for measuring and calibrating cluster masses is required and lensing calibrations are expected to be the most unbiased \citep[e.g.,][]{Becker2011}.

Planned CMB experiments like Advanced ACT, SPT-3G, Simons Array, Simons Observatory (SO) and CMB Stage-4 will produce catalogs of tens of thousands of galaxy clusters. Overlap with an optical survey like LSST will allow for precise measurements of the shapes of background galaxies behind these galaxy clusters, which will provide accurate mass calibrations for most of the galaxy clusters at low and intermediate redshifts. Another promising and independent way to calibrate the observable--mass relation is to use the so-called {\it CMB halo lensing} \citep{SZ2000,Dod2004,HK2004,HDV2007}, lensing of the CMB itself \citep{LA2017} by dark matter halos. This technique is viable for clusters at any redshift and has completely different systematics from optical weak lensing. The lensing signal from dark matter halos has only recently been detected \citep{Mat2015,Baxter2015,PlnkSZCos2015,GP2017,Baxter2017} and has already been used as a mass calibrator in an tSZ cosmological analysis \citep{PlnkSZCos2015}.

In this paper we forecast the constraints that CMB Stage-4 can achieve on cosmological parameters using tSZ cluster counts. In the forecasts we include the external calibrations of the tSZ observable--mass relation for clusters from optical weak-lensing observations using experiments like LSST and internal CMB Halo lensing calibration from CMB Stage-4. The paper is structured as follows: in Section~\ref{sec:exp} we describe the assumptions about the experimental setup of CMB Stage-4 and its variations that we compare. Section~\ref{sec:meth} describes our methodology for modeling cluster detection and cluster abundance. In Section~\ref{sec:owl}, we present how we forecast the ability of an LSST-like experiment to calibrate the tSZ scaling relation and in Section~\ref{sec:cmbhalo}, we describe how CMB lensing does the same. Section~\ref{sec:res} presents our Fisher forecasting assumptions and the cosmological models considered. We conclude with a discussion in Section~\ref{sec:con}.

\section{Experimental setup}
\label{sec:exp}
We consider an experimental configuration consisting of a single large telescope with seven band-passes shown in Table~\ref{tab:cmbexp}. Our baseline configuration has a white noise level of $s_{\nu,w}=1.5\mu K'$ in the 150 GHz and 90 GHz channels. We vary the beam full-width half-maximum (FWHM) $\theta_{b}$ in the 150 GHz channel from 1 arcminute to 3 arcminutes, scaling the beam FWHM in the other channels $\nu$ assuming $\theta_b\propto 1/\nu$. The noise sensitivities are assumed to correspond to a fraction of sky observed $f_{\mathrm{sky}}=0.4$.

In addition to instrumental white noise, we include the effect of atmospheric noise for a ground-based experiment, parameterized through a knee multipole $\ell_{\mathrm{knee}}$ and tilt $\alpha$,

\begin{equation}
N_\nu(\ell) = s_{\nu,w}^2\left(1+\left(\frac{\ell}{\ell_{\mathrm{knee}}}\right)^{\alpha}\right).
\end{equation}

Atmospheric noise can have a considerable impact on the number of clusters detected and consequently on cosmological constraints. In our fiducial analysis, we will assume an $\ell_{\mathrm{knee}}$ of 3500 in temperature and 300 in polarization, and an ${\alpha}$ of -4.5 in temperature and -3.5 in polarization in every frequency bandpass \citep{Louis2017}. In reality, these parameters may depend on the aperture size of the telescope (amongst other experimental variables) and will vary between bandpasses. Our fiducial values are motivated by the performances of past and ongoing ground-based CMB experiments. A detailed analysis of the dependence of $\ell_{\mathrm{knee}}$ and ${\alpha}$ on aperture size and frequency of observation is beyond the scope of this work. We do however undertake a study of the effect of $\ell_{\mathrm{knee}}$ and ${\alpha}$ on the number of clusters detected in Section III.

As described in Section~\ref{sec:meth}, each assumed experimental configuration predicts a certain number of tSZ cluster detections as a function of mass, redshift and signal-to-noise. For the sample of tSZ clusters selected this way, we obtain lensing mass calibration either internally using CMB lensing (see Section~\ref{sec:cmbhalo}), or externally from an LSST-like optical weak lensing survey configuration (see Section~\ref{sec:owl}). The internal calibration is done either on both temperature and polarization data (T+P) from the 150 GHz channel, or on polarization only (P-only). The optical lensing calibration is done either for clusters with redshifts $0<z<1$ or for $0<z<2$. We assume that the optical lensing survey provides brightest central galaxies (BCG) which are used as centroids for stacking the optical and CMB lensing signals. We therefore assume that mis-centering of the stack and the true mass centroid can be assumed to be negligible (compared to the beam size) for clusters with $z<2$. For CMB lensing mass calibration of $2<z<3$ clusters, we do not assume the availability of BCG centers and marginalize over mis-centering effects as described in Section~\ref{sec:cmbhalo}.

The optical survey is also assumed to provide photometric redshifts for at least some member galaxies of each tSZ detected cluster. These redshifts are not required to be very precise since they are needed only for coarse binning of the clusters in redshift. Any tSZ detected cluster that cannot be associated with any possible member galaxies in the optical survey can fairly confidently be assigned to the $2<z<3$ redshift bin that is calibrated using CMB lensing. 

\begin{table*}
  \caption[CMB Experimental setup]{The experimental configurations considered for CMB Stage-4. The frequency band-passes and map sensitivities are fixed at the values in the first two columns, while beam FWHMs are varied for five configurations. }
  \label{tab:cmbexp}
  \begin{center}
   \leavevmode
   \begin{tabular}{c|c|c|c|c|c|c} 
     \hline \hline
     
& & \multicolumn{5}{c}{Beam (arcminutes)} \\
          Frequency (GHz) & Noise [$\mu K'$] &  CMB-S4-3.0$'$ & CMB-S4-2.5$'$ &  CMB-S4-2.0$'$ &  CMB-S4-1.5$'$ &  CMB-S4-1.0$'$ \\
     \hline
21 & 7.9& 21.4 & 17.9 & 14.3 & 10.7 & 7.1 \\ 
29 & 5.6 &15.5 & 12.9 & 10.3 & 7.8 & 5.2 \\ 
40 & 5.4 & 11.2 & 9.4 & 7.5 & 5.6 & 3.8 \\ 
95 & 1.5 &4.7 & 4.0 & 3.2 & 2.4 & 1.6 \\ 
150 & 1.5 & 3.0 & 2.5 & 2.0 & 1.5 & 1.0 \\ 
220  & 5.2 & 2.0 & 1.7 & 1.4 & 1.0 & 0.7 \\ 
270  & 9.0 & 1.7 & 1.4 & 1.1 & 0.8 & 0.6 \\
\hline
   \end{tabular}
  \end{center}
  \begin{quote}
    \noindent 
We emphasize that this is a straw-person experimental design for CMB Stage-4, for example the frequency bands for CMB Stage-4 have not yet been determined. These particular bands were chosen to cover the main atmospheric windows around the peak CMB and tSZ sensitivity with extra high and low frequency bands for potential foreground cleaning. The distribution of detectors weights among bands reflects a rough optimization for CMB and tSZ signal and assuming some level of foreground subtraction. In this work we focused on the aperture size of CMB Stage-4 and a full optimization of the frequency bands, noise levels, and aperture sizes is beyond the scope of this work. In particular frequency bands optimization requires simulations that include correlated sources of noise like the analyses in \citet{Melin2017}.
 \end{quote}

\end{table*}

\section{Methodology}
\label{sec:meth}
The thermal SZ (tSZ) signal is the observable cluster property that we model onto the theoretical predictions for the abundance of clusters. We use an analytic model for these tSZ selected clusters that accounts for measurement uncertainties in mass calibration and integrated Compton-$y$ signal. The spectral distortion caused by the tSZ in the observed CMB temperature is a function of frequency $\nu$\footnote{for a theoretical experiment $\nu$ represents the central frequency of a given frequency band} and the Compton-$y$ parameter ($y$): 

\begin{equation}
\frac{\Delta T(\nu)}{T_\rmn{CMB}} = f_\nu y,
\label{eq:delt_tsz}
\end{equation}
here $f_\nu = x\,\rmn{coth}(x/2) - 4$, where $x = h\nu / (k T_\rmn{CMB})$, $h$ is the Planck constant, and $k$ is the Boltzmann constant. Note that we neglected relativistic corrections to the tSZ spectral function $f_\nu$ \citep[e.g.,][]{Nozawaetal2006,Chluba2012}. As shown in Equation~\ref{eq:delt_tsz}, the amplitude of the tSZ spectral distortion is directly proportional to $y$, which is defined as, 

\begin{equation}
y = \frac{\st}{\me c^2} \int \nume kT_\rmn{e} \dd l.
\label{eq:y}
\end{equation}

\noindent Here the physical constants $c$, $\me$, and $\st$ correspond to the speed of light, electron mass, and Thompson cross-section, respectively. The physical properties of the free electron that scatter the CMB photons are: $\nume$ the electron number density and $T_\rmn{e}$ is the electron temperature. Equation~\ref{eq:y} is integral along the line-of-sight, $\dd l$. For a given spherical pressure profile, $P_\rmn{e} (r)= \nume (r) kT_\rmn{e} (r)$ the $y$ signal for a cluster projected on the sky is,

\begin{equation}
y(\tht) = \frac{\st}{\me c^2}  \int P_\rmn{e} \left(\sqrt{l^2 + d^2_A(z)|\tht|^2 } \right) \dd l.
\label{eq:yproj}
\end{equation}

\noindent Here $r^2 = l^2+d^2_A(z)|\tht|^2$, $d_A(z)$ is the angular diameter distance to redshift $z$, and $\tht$ is the 2D angular coordinate on the sky.

For the shape of $y(\tht)$, we choose the pressure profile from \citep{Arnd2010}, which was used in the Planck cluster analysis \citep[e.g.,][]{PlnkSZCos2015}. The parametric form for the profile is a generalized Navarro-Frenk-White profile \citep{Zhao1996}, 

\begin{equation}
P_\rmn{e}(x) = P_0 \left(c\, x\right)^{- \gamma} \left[1 +  \left(c\, x\right)^{ \alpha} \right]^{\frac{\gamma - \beta}{\alpha}}, 
\end{equation}
where the $x = r/R_{500 \rho_c}$, and the parameters of the profile have the values, $P_0 = 8.403$, $c = 1.156$, $\alpha = 1.062$, $\gamma = 0.3292$, and $\beta = 5.4807$. Additionally, we choose the filter scale for each cluster, $\Theta_{500 \rho_c}(z)$, such that $\Theta_{500 \rho_c}(z) = R_{500\rho_c} / D_A(z)$.

\subsection{Cluster detection}

We find tSZ clusters using a matched filter technique that exploits the unique spectral distortion of the tSZ effect \citet{Herranz2002,JB2006}. We model the maps of the millimeter sky, $ \mathcal{M}(\tht)$, as:

\begin{equation}
\mathcal{M}_\nu(\tht) = Y_0 f_\nu g(\tht) + N_\nu(\tht)
\end{equation}

\noindent here, $Y_0$ is the amplitude of the tSZ signal for a given halo, $g(\tht)$ is the normalized projected $y$ profile, $g(\tht) = y(\tht) / Y_0$, and $N_\nu(\tht)$ is the noise when searching for a tSZ signal. Here the noise is a function of $\nu$ and includes instrumental noise, atmosphere (described in Section~\ref{sec:exp}), primary CMB, and other secondary sources.

The estimator we use to measure $Y_0$ is a matched filter that is designed to minimize the variance across a given set of frequency bands for an assumed $y(\tht)$ profile

\begin{equation}
\hat{Y_0} = \int F_\nu(\tht)^T M_\nu(\tht) \dd \tht .
\label{eq:mf}
\end{equation}
Here we sum over $\nu$ and $F_\nu(\tht)$ is an unbiased, real-space matched filter that minimizes the variance. In Fourier space this matched filter has the form,

\begin{equation}
F_\nu(\ell) = \sigma_N^2 \left[{C}_{N,\nu \nu'}(\ell)\right]^{-1} f_{\nu'} \tilde{g}(\ell).
\end{equation}
Here $\tilde{g}(\ell)$ is the Fourier transform of the normalized projected $y$ profile, $\sigma_N^2$ is the variance, and $C_{N,\nu \nu'}(\ell)$ is the covariance matrix of the noise power spectrum. Note that the Fourier transform of $F_\nu(\ell)$ is $F_\nu(\tht)$. The variance is defined as 

\begin{equation}
\sigma_N^2 = 2 \pi \int |\tilde{g}(\ell)|^2 f_\nu^T \left[{C}_{N,\nu \nu'}(\ell)\right]^{-1} f_{\nu'}\, \ell \, \dd \ell,
\label{eq:erry}
\end{equation}
and the noise covariance matrix is defined as,
\begin{eqnarray}
{C}_{N,\nu \nu'} (\ell) &=& {C}_{\mathrm{CMB},\nu \nu'} (\ell) + {C}_{\mathrm{sec},\nu \nu'} (\ell) \nonumber \\
&+& \left(\frac{N_\nu(\ell)}{B_{\nu}(\ell)^2 }\right)\delta_{\nu \nu'}.
\end{eqnarray}
The components of the noise covariance matrix are the CMB cross-power spectra ${C}_{\mathrm{CMB},\nu \nu'} (\ell)$, the secondary cross power spectra ${C}_{\mathrm{sec},\nu \nu'} (\ell)$, and the de-beamed noise $ N_\nu(\ell) / B_{\nu}(\ell)^2$ that only contributes to the diagonal, where $B_{\nu}(\ell)$ is the Fourier transform of the beam, which we assume to be Gaussian. The FWHM of the beams for corresponding frequency bands are shown in Table~\ref{tab:cmbexp}. The CMB secondary anisotropies that we include are, radio point sources (Poisson term), the cosmic infrared background (CIB, both Poisson and clustered terms), kinetic Sunyaev-Zel'dovich, unresolved tSZ, and the tSZ-CIB cross-correlation term \citep[][]{Addison2012}. For the unresolved tSZ contribution, we estimate that half of the total auto-spectrum power is coming from clusters with masses $\sim 10^{14} M_\odot$ \citep[e.g.,][]{KS2002,Trac2011,BBPS2}, that will be detected. Therefore, we removed the contribution from these clusters to the auto-spectrum power for the purposes of additional secondary anisotropy noise. We use the functional forms and parameters for these secondary anisotropies presented in \citet{Dunkley2013}.

\begin{figure}[t]
\includegraphics[width=0.95\columnwidth]{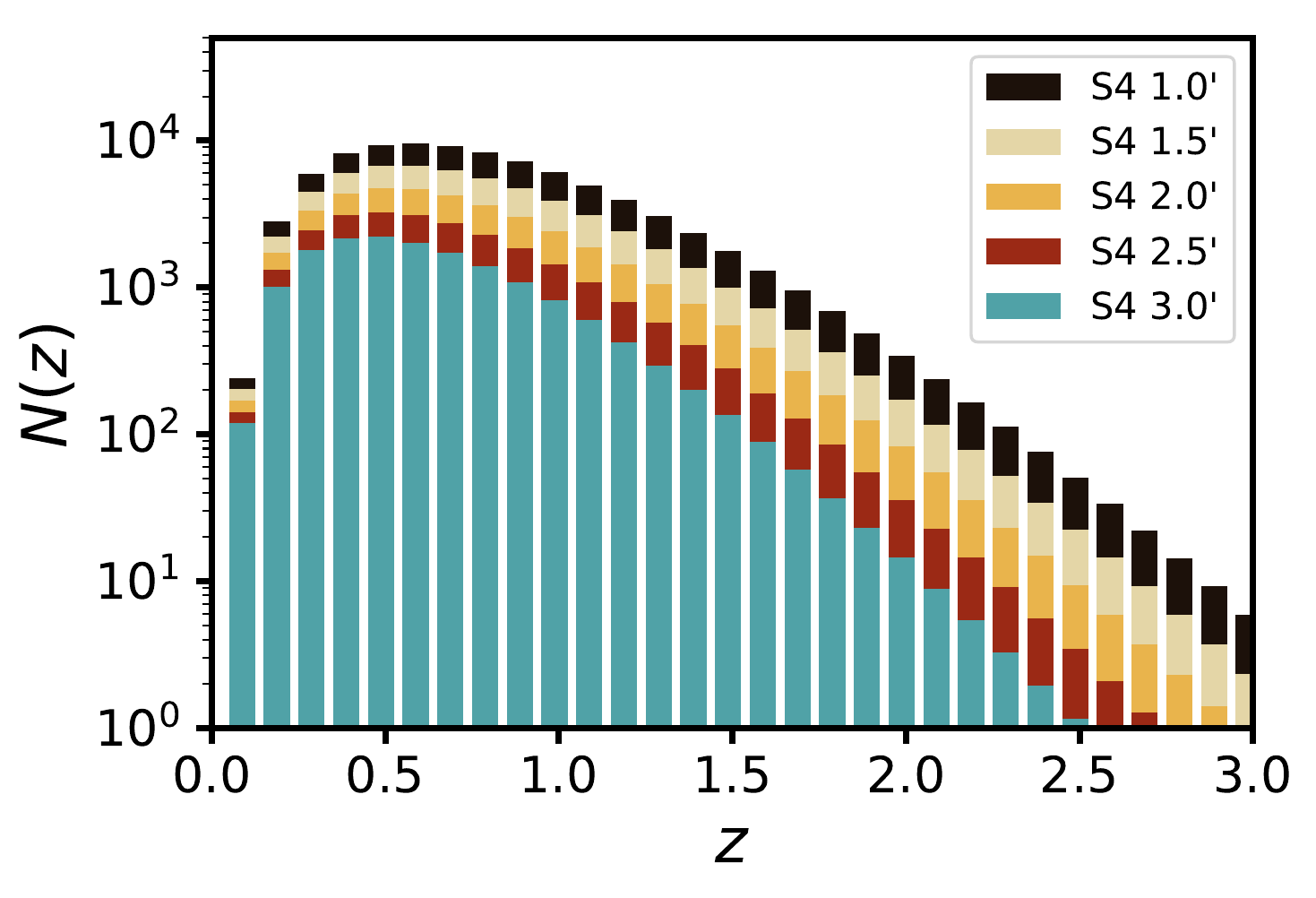}
\includegraphics[width=0.95\columnwidth]{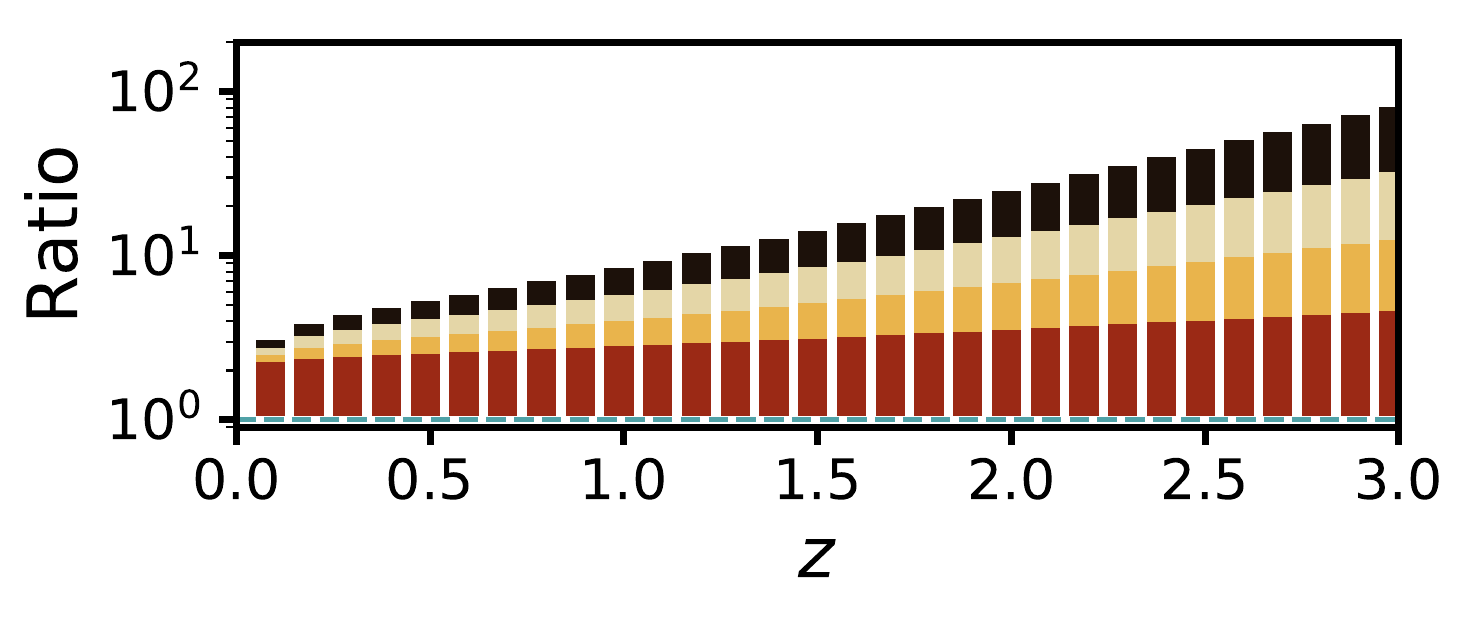}
\caption{{\it Top:} The number of clusters that can be detected through tSZ emission by CMB Stage-4 in each redshift bin of width $\Delta z = 0.1$. The different colors correspond to different beam FWHMs in the 150 GHz channel. {\it Bottom: } The ratio of the number of clusters at various resolutions to that for a telescope with $3'$ resolution at 150 GHz.}
\label{fig:dn_dz}
\end{figure}

\begin{figure}[t]
\includegraphics[width=0.95\columnwidth]{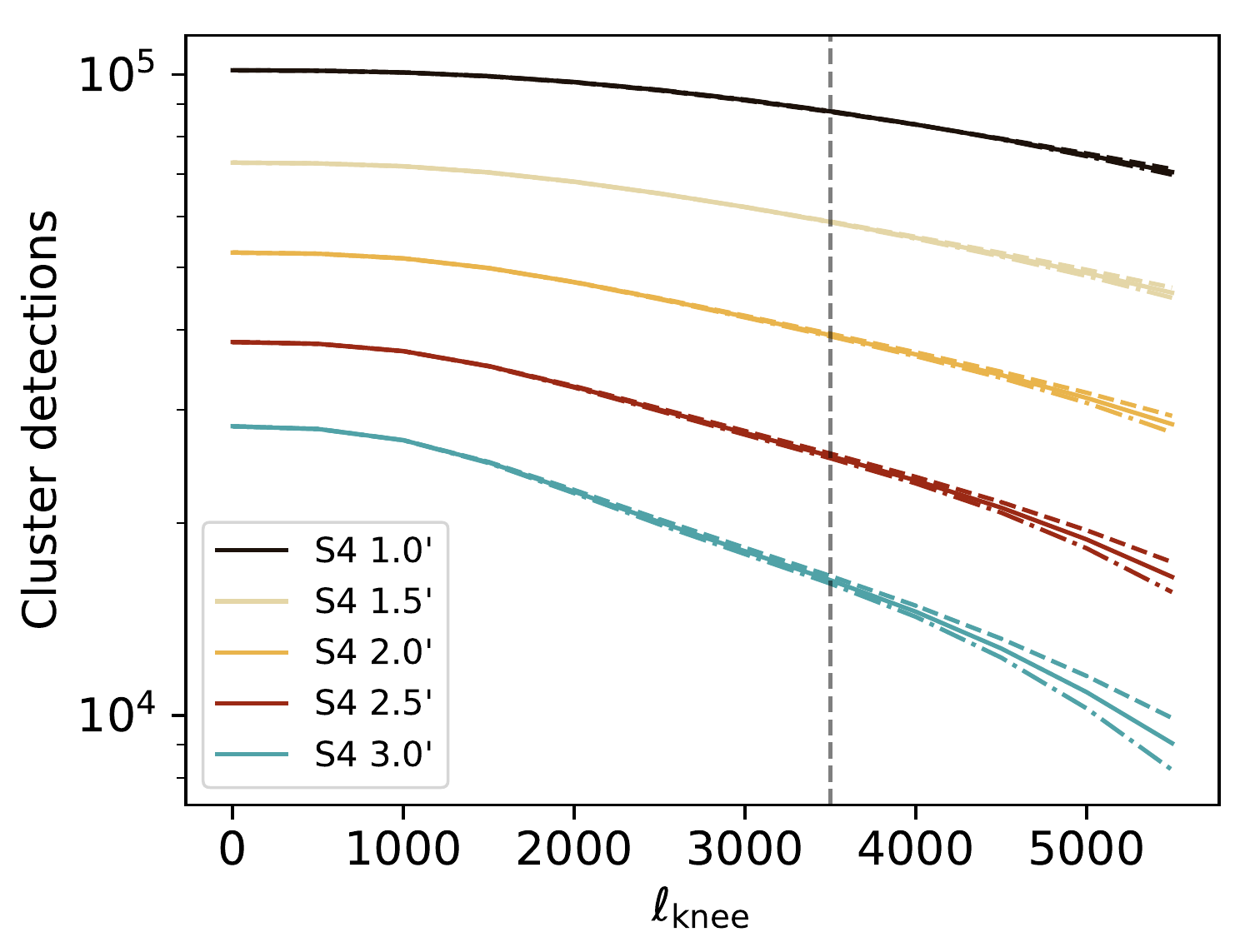}
\caption{The number of cluster detections as a function of the $\ell_{\mathrm{knee}}$ atmospheric noise parameter for each resolution considered and for three different values of the $\alpha$ atmospheric noise parameter. The solid lines correspond to $\alpha=-4.5$, dashed to $\alpha=-4$ and dot-dashed to $\alpha=-5$. The vertical dashed line corresponds to the $\ell_{\mathrm{knee}}$ used in the rest of the analysis.}
\label{fig:atm}
\end{figure}

\subsection{tSZ cluster abundances}

For simplicity the abundance of tSZ clusters is modeled as the number of clusters ($N$) observed in bins of lensing mass calibration ($M_\rmn{L}$), tSZ signal-to-noise ($q_\rmn{Y}$) from the matched filter, and redshift ($z$):

\begin{eqnarray}
\frac{N (M_\rmn{L},q_\rmn{Y},z)}{\Delta M_\rmn{L} \Delta q_\rmn{Y} \Delta z} &=& \int \frac{\dd^2 N}{\dd z \dd M} P(M_\rmn{L},q_\rmn{Y} | Y , M) \nonumber \\
& & \N(\rmn{log} Y|\rmn{log}\bar{Y},\sigma_{\rmn{log}Y}  ) \dd M \dd Y.
\label{eq:abund}
\end{eqnarray}
Here $P(M_\rmn{L},q_\rmn{Y} | Y , M)$ is the probability distribution function of $M_\rmn{L}$ and $q_\rmn{Y}$ given the integrated Compton-y, $Y$, and halo mass, $M$ (see Equation~\ref{eq:pq}), $\frac{\dd^2N}{\dd M\dd z}$ is the differential number of clusters with respect to $M$ and $z$ (see Equation~\ref{eq:dn}), and $\N(\rmn{log} Y|\rmn{log}\bar{Y},\sigma_{\rmn{log}Y})$ is a lognormal distribution of $Y$ given the mean integrated Compton-y ($\bar{Y}$, see Equation~\ref{eq:ym}) and the intrinsic scatter ($\sigma_{\rmn{log}Y}$ see Equation~\ref{eq:sigy}). 

We model the probability distribution function 
of $M_\rmn{L}$ and $q_\rmn{Y}$ given $Y$ and $M$  as two independent normal distributions ($\N$), 
\begin{eqnarray}
\label{eq:pq}
P(M_\rmn{L},q_\rmn{Y} | Y , M) &=& \N (q_\rmn{Y} | Y/\sigma_Y , 1)  \nonumber \\
&\,& \N(M_\rmn{L} b_\rmn{L}|M,\sigma_M).
\end{eqnarray}
The $Y$ measurement errors, $\sigma_Y$, is determined from the matched filter (see Equation~\ref{eq:erry}) and $M_\rmn{L}$ measurement errors, $\sigma_M$, comes from either the optical weak-lensing or CMB halo-lensing mass calibration (see Sections~\ref{sec:owl} and~\ref{sec:cmbhalo} for details). The parameter $b_\rmn{L}$ is set to $b_\rmn{L}=1$ for the main analysis but is allowed to vary with a 1\% Gaussian prior when we wish to explore the imposition of a 1\% systematic floor on the mass calibration. We apply $b_\rmn{L}$ to $M_\rmn{L}$ and not $\sigma_M$ since we want to impose a systematic floor that is independent of $\sigma_M$ and is irreducible.

The differential numbers counts can be further deconstructed into
\begin{equation}
\label{eq:dn}
\frac{\dd^2N}{\dd M\dd z} = \frac{\dd V}{\dd z \dd \Omega} n(M,z). 
\end{equation}
Here the volume element is $\frac{\dd V}{\dd z \dd \Omega}$ and for the mass function, $n(M,z)$, we use \citet{Tink2008} while accounting for the neutrino suppression of power in $\Omega_M$ after recombination, where $\Omega_M = \Omega_b + \Omega_\rmn{CDM}$ and not $\Omega_\nu$ \citep[e.g.,][]{IT2012,Cost2013,Cast2014,Cast2015}.

Following the theoretically motivated self-similar evolution of halos \citep{Kaiser1986} we model the $Y$--$M$ scaling relation as a power-law that is a function of halo mass and redshift, 

\begin{eqnarray}
\label{eq:ym}
\bar{Y} (M,z) &=& Y_\star \left[\frac{(1-b)M_{500\rho_c}}{M_\star}\right]^{\alpha_Y} e^{\beta_Y \rmn{log}^2(M_{500\rho_{c}}/M_\star)} 
\nonumber \\
&\,&E^{2/3}(z) (1 + z)^{\gamma_Y} \left[\frac{D_A(z)}{100 \rmn{Mpc}/h}\right]^{-2}.
\end{eqnarray}
Where $Y_\star = 2.42 \times 10^{-10}$ is a constant, the pivot mass is $M_\star = 10^{14}\,M_\odot$/h, $1 - b$ is the mass bias correction, $\alpha_Y$ is the first order power-law mass dependence, $\beta_Y$ is the second order power-law mass dependence, and $\gamma_Y$ is an additional redshift dependence beyond the expected self-similar scaling. The functions $E(z)$ and $D_A(z)$ are the Hubble function and the angular diameter distance, respectively. The fiducial values for the scaling relation parameters in this model are $\{1-b,\alpha_Y,\beta_Y,\gamma_Y\} = \{0.8,1.79,0,0\}$. We model the scatter in this scaling relation as, 

\begin{equation}
\label{eq:sigy}
\sigma_{\rmn{log}Y} (M,z) = \sigma_{\rmn{log}Y,0} \left[\frac{M_{500\rho_c}}{M_\star}\right]^{\alpha_{\sigma}} (1 + z)^{\gamma_{\sigma}},
\end{equation}
where $\sigma_{\rmn{log}Y,0}$ is the fiducial scatter and the smooth power-law mass and redshift dependence of the scatter are $\alpha_{\sigma}$ and $\gamma_{\sigma}$, respectively. We choose the fiducial values for these scatter parameters to be $\{\sigma_{\rmn{log}Y,0},\alpha_{\sigma},\gamma_{\sigma}\} = \{0.127,0,0\}$. All fiducial parameter values and their associated step sizes for our Fisher analyses are shown in Table II.

In Figure~\ref{fig:dn_dz} we show the number of expected clusters that would be detected as a function of redshift for various choices of aperture size for CMB Stage-4. The bottom panel illustrates the ratio of clusters compared to the most pessimistic design of the CMB Stage-4 experiment (3 arcminute aperture at 150 GHz) for the purposes of tSZ and secondary anisotropy science. As a function of increasing aperture the gain in detected clusters increases strongly with redshift. Between $z = 2 - 3$, we find increases on the order of hundreds when comparing a 1 arcminute to a 3 arcminute aperture. This redshift range of $z = 2 - 3$ is a new frontier for clusters and proto-clusters science, with only a few heterogeneously detected clusters and proto-clusters within this redshift range \citep[][and references therein]{Over2016}. If these clusters contain hot gas, as we have assumed, CMB Stage-4 will find them and produce a legacy catalog of uniformly selected, high-$z$  tSZ clusters that will be ideal to study galaxy formation in high redshift, dense environments.

Atmospheric noise primarily affects large scales in the CMB, but with a sufficiently high knee multipole its effects can degrade scales relevant for cluster finding. We explore the effect of atmospheric noise in Figure \ref{fig:atm}. There is a strong dependence on $\ell_{\mathrm{knee}}$, e.g., $\ell_{\mathrm{knee}}=5500$ for the 1-arcminute configuration corresponds to detecting 30\% fewer clusters than if there were no atmospheric noise ($\ell_{\mathrm{knee}}=0$).

\begin{figure}[t]
\includegraphics[width=0.95\columnwidth]{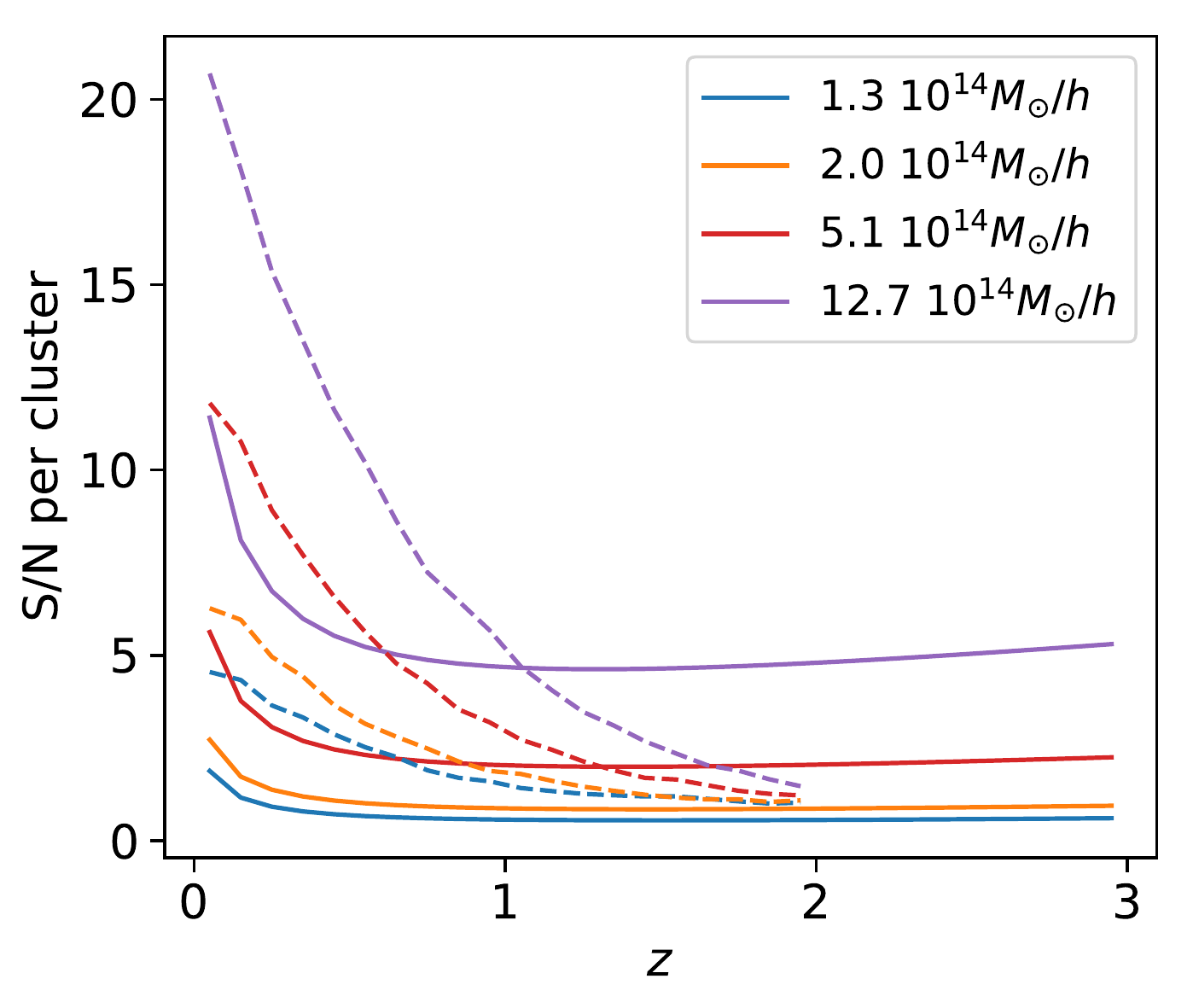}
\caption{The dependence of signal-to-noise of the lensing signal per cluster as a function of redshift for an LSST-like optical survey (dashed lines) and internal CMB lensing calibration using both temperature and polarization (solid lines) with a telescope that has beam FWHM of $1'$ at 150 GHz. The different colors correspond to specific $M_{500\rho_c}$ masses of clusters. The increase in S/N at low redshift is partly due to clusters becoming more concentrated at lower redshifts, since we assume a $c_{500}$ dictated by the relation in \citet{Duffy:2008}. For clusters with $M_{500\rho_c} \gtrsim 5\times 10^{14} M_\odot$ CMB lensing offers higher S/N per cluster at some redshift $z>1$ and for all clusters above $z=2$ where the ability to measure shapes of background galaxies degrades quickly.}
\label{fig:cmbVoptical}
\end{figure}

\section{Optical Weak Lensing}
\label{sec:owl}
We model the optical lensing signal as follows. For a given cluster mass $M_{\rm 500c}$, we compute concentration $c_{\rm 500c}$ using the the concentration-mass relation derived in \cite{Duffy:2008}. We then convert $(M_{\rm 500c}, c_{\rm 500c})$ to $(M_{\rm 200m}, c_{\rm 200m})$ assuming the Navarro-Frenk-White \citep[NFW,][]{Navarro:1997} profile. We assume the optical weak lensing signal at redshift $z_l$ is measured in terms of excess surface density (ESD);
\begin{eqnarray}
\Delta\Sigma(R;M_{\rm 200m}, c_{\rm 200m}, z_l) &=&
\langle \Sigma(<R;M_{\rm 200m}, c_{\rm 200m}, z_l) \rangle \nonumber\\ &&- \Sigma(R;M_{\rm 200m}, z_l),
\end{eqnarray}
where $R$ is the galaxy-centric transverse comoving distance, $\Sigma(R)$ is the projected matter density profile along the line-of-sight, and $\langle \Sigma(R) \rangle$ is the projected matter density profile averaged over distance $R$. We employ the following halo model for ESD;
\begin{eqnarray}
\Delta\Sigma(R;M_{\rm 200m}, c_{\rm 200m}, z_l) &=& \Delta\Sigma^{\rm NFW}(R;M_{\rm 200m}, c_{\rm 200m}, z_l) \nonumber\\ 
&&+ \Delta\Sigma^{\rm 2h}(R;M_{\rm 200m}, z_l),
\end{eqnarray}
where $\Delta\Sigma^{\rm NFW}(R)$ is a smoothly-truncated version of the NFW profile proposed in \cite{Baltz:2009} with the dimensionless smoothing radius $\tau\equiv r_t/r_{\rm 200m}=2.6$, which is converted from $\tau$ defined against virial radius in \cite{Oguri:2011}. We do not fit for the 2-halo term $\Delta\Sigma^{\rm 2h}(R)$, since we restrict the our analyses to the regime where $\Delta\Sigma^{\rm NFW}(R)$ dominates.



The shape noise of a given radial bin $R$ with sources at redshift $z_s$ is estimated as
\begin{equation}
\label{eq:shape_noise_z}
\sigma^2\left(\Delta\Sigma(R,z_l);z_s\right)=\frac{\sigma_g^2 \Sigma_{\rm cr}^2(z_l, z_s)} {A \, n_g(z_s)},
\end{equation}
where $\sigma_g^2$ is the RMS of intrinsic ellipticity\footnote{In this equation, ellipticity is defined in terms of {\it shear}, i.e., $g=(a-b)/(a+b)$, where $a$ and $b$ is the major and minor axis, respectively}, $A$ is the area of the radial bin, and $n_g(z_s)$ is the number density of source galaxies.
The critical surface mass density $\Sigma_{\rm cr}$ is defined as
\begin{equation}
\Sigma_{\rm cr}(z_l,z_s) = \frac{c^2}{4\pi G} \frac{D_A(z_s)}{(1+z_l)^2D_A(z_l)D_A(z_l,z_s)},
\end{equation}
where $D_A$ is the angular diameter distance of the lens-source system and $(1+z_l)^{-2}$ comes from our use of comoving coordinates \citep{Mandelbaum:2006}. The total shape noise is estimated by assuming Eq.~\eqref{eq:shape_noise_z} forms the inverse variance weight for given  redshift of the lens;
\begin{equation}
\sigma^2\left(\Delta\Sigma(R,z_l)\right) = \left[ \int_{z_l}^{\infty} \sigma^{-2}\left(\Delta\Sigma(R,z_l);z_s\right) \dd z_s\right]^{-1}.
\end{equation}

We compute the shape noise by asserting that there will be HSC-like survey over the entire $f_{sky}$ of the survey. This is not unreasonable considering LSST will be available over a large area if not all of the CMB Stage-4 survey area and the parameters of HSC are conservative compared to LSST. For the HSC survey we assumed a source background 20 galaxies per square arcminute with the $\dd N_g / \dd z$ from \cite{Oguri:2011};
\begin{equation}
\frac{\dd N_g}{\dd z} = \frac{z^2}{2z_0^3}\exp\left(-\frac{z_s}{z_0}\right),
\end{equation}
where $z_0=1/3$ that corresponds to the mean redshift $z_m=1$.

We fit $\Delta \Sigma (r)$ given the derived shape noise errors with an NFW profile over the radial range where the 1-halo term is determined using Markov Chain Monte Carlo  (MCMC) \citep{emcee} assuming a fixed concentration mass relation \citep{Duffy:2008}. The radial range we use for the fit is $\sim 0.1-4$ in comoving Mpc. We use the width of the inferred weak lensing mass distribution as our weak lensing mass error and take the ratio of this over the median inferred mass as the percent weak lensing mass error $\Delta M / M$. Figure~\ref{fig:cmbVoptical} illustrates the S/N per cluster as a function of redshift.

We caution that systematic errors are not taken into account when we forecast the errors bars on $\Delta \Sigma (r)$. We expect systematic uncertainties to be increasingly important as the redshift of the clusters increases due to photometric redshift and shape measurement biases \citep[e.g.,][]{Jarvis2016,Tanaka2017}. For this reason, we compare constraints from clusters with redshifts $0<z<1$ and $0<z<2$. It is possible that at higher redshifts such systematic uncertainties will be larger than 1\%, in this regime CMB halo lensing will become important.

\section{CMB Halo Lensing}
\label{sec:cmbhalo}

The CMB is lensed by all structure since recombination and hence can in principle be used as a source for any galaxy cluster. In contrast to optical lensing where the sources are distributed in a wide range of uncertain redshifts behind the cluster, the CMB source plane is fixed at a relatively thin and well-measured slice at $z=1090$. In addition, the CMB is a diffuse field whose unlensed statistics are well captured by a Gaussian random field specified through a power spectrum. The effect of lensing is to to couple previously independent harmonic modes of the CMB temperature (T) and polarization fields (curl-free E and curl-like B). This insight allows one to write a quadratic estimator that sums over pairs of CMB modes optimally to reconstruct the projected lensing potential at any given mode \cite{HDV2007}.

The quadratic estimator requires a pair of maps, the first `leg' of the pair effectively serving as a measure of the background gradient at the location of the cluster, and the second leg capturing information about the small-scale fluctuations induced by lensing by the cluster. For example, in the temperature-only estimator combination (TT), the quadratic estimator can be written as \cite{HDV2007}
$$
\kappa(\boldsymbol{\theta}) =\mathcal{F}^{-1}\left\{A(\boldsymbol{L}) \mathcal{F}\left\{ \nabla . \left[\boldsymbol{\nabla}T(\boldsymbol{\theta}) T(\boldsymbol{\theta}) \right]\right\}\right\}
$$
where $\boldsymbol{\nabla}T(\boldsymbol{\theta}) $ is the low-pass filtered gradient of the temperature map as a function of angle on the sky $\boldsymbol{\theta}$, $T(\boldsymbol{\theta})$ is the high-pass filtered temperature map, $A(\boldsymbol{L})$ is a normalization in Fourier-space that ensures this estimator is unbiased as a function of angular wave-number $\boldsymbol{L}$, and $\mathcal{F}$ and $\mathcal{F}^{-1}$ represent 2D Fourier and inverse-Fourier transforms respectively.

  We use the estimator combinations TT, TE, EE, EB and TB, where the first leg in the pair is used in gradient estimation and the second leg is used for small-scale fluctuations. The noise per mode in each estimator is $N^{\kappa\kappa}_{L}=L^2A_L/4$ where the estimator normalization $A_L$ is given by Equation 19 in \citet{HDV2007}. We consider minimum variance combinations of either all the above estimators (T+P) or `Polarization Only' (P only), i.e., EE and EB. We do not account for the covariance between these estimators since in most cases either TT or EB dominates, and the covariance between TT and EB is zero. We then calculate the total signal and variance of the lensing convergence $\kappa$ measured within a radius of $5\theta_{500}$ when using a matched filter designed to optimally measure the lensing signal.

In this forecast analysis, we only use information for CMB lensing calibration from the CMB Stage-4 150 GHz channel, but use multi-frequency information from the Planck satellite experiment. Planck has effectively imaged CMB temperature modes at scales $\ell<2000$ to nearly the cosmic variance limit (even if considering a foreground cleaned map like SMICA \citep{SMICA} or LGMCA \citep{LGMCA1,LGMCA2}). The CMB halo lensing signal can be contaminated by both noise and systematic biases from cluster foregrounds such as tSZ and CIB. This contamination is to some extent mitigated by the imposition of a low-pass filter $\ell<2000$ on the gradient leg. However, the systematic bias from cluster contaminants can be nearly eliminated by requiring that the gradient leg be foreground free. Since this leg only requires modes $\ell<2000$, we lose almost no signal-to-noise if we use Planck foreground cleaned maps in the temperature gradient leg of the TT, TE and TB estimators while eliminating the largest source of potential bias. The rest of this analysis assumes Planck beam (5 arcminutes) and white noise of 42 $\mu K$-arcmin for the first leg of TT, TE, TB (corresponding roughly to the level in a foreground cleaned map) and CMB Stage-4 150 GHz (Table~\ref{tab:cmbexp}) in all the other legs of the estimator combinations.

\begin{figure}[t]
\includegraphics[width=0.95\columnwidth]{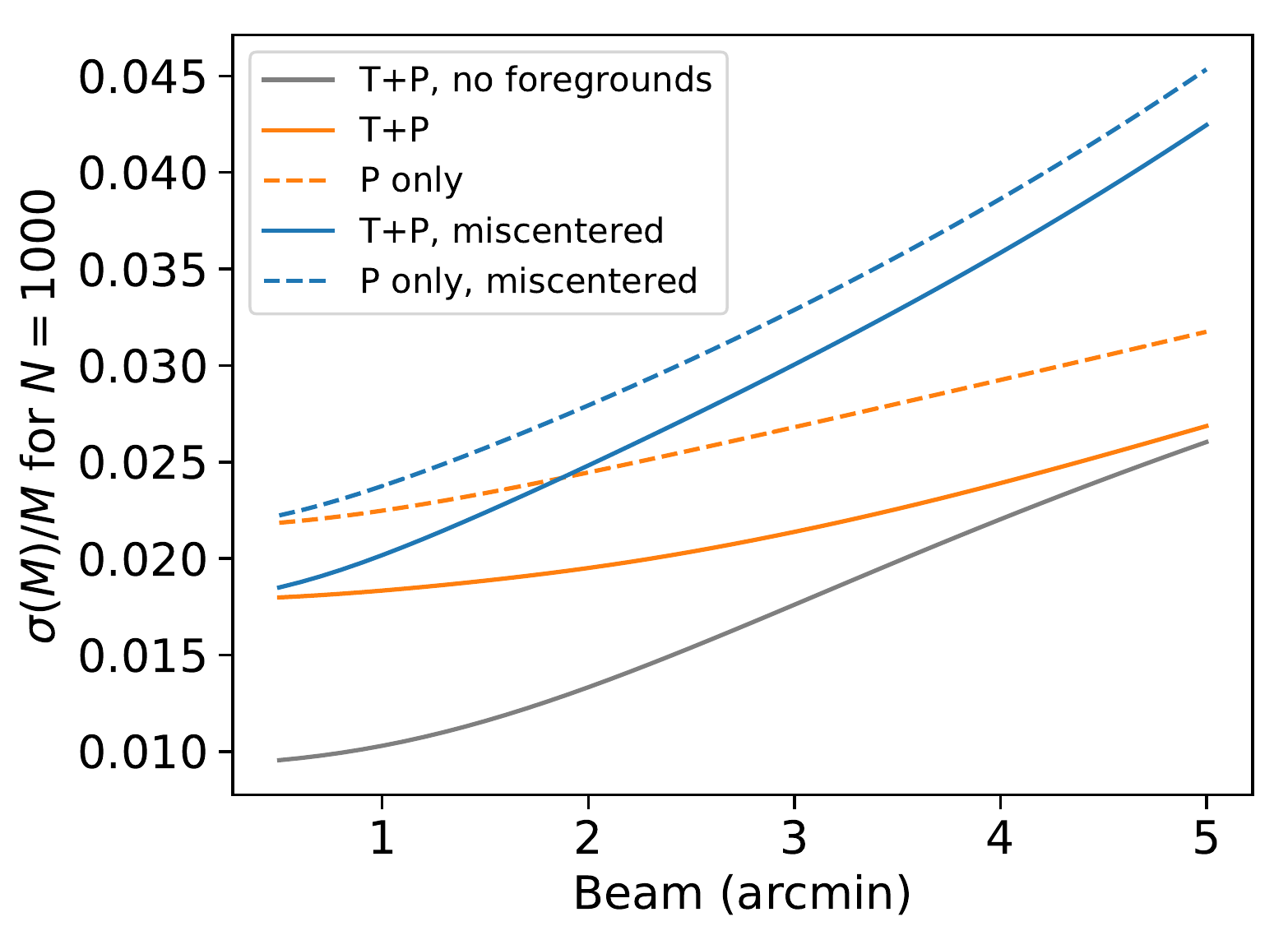}
\caption{The mass sensitivity (uncertainty on $M_{500}$ over $M_{500}$) for $N=1000$ clusters with $M_{500}=2\times 10^{14}M_{\odot}/h$, $z=0.7$, $c_{500}=1.18$ as a function of the beam FWHM at 150 GHz when masses are estimated using a matched filter on lensing reconstructions from CMB Stage-4. The solid curves use both CMB temperature and polarization data while the dashed curves only use polarization data. The gray curve assumes that there is no noise from astrophysical foregrounds in the temperature maps. Polarization maps are always assumed to be foreground-free. The blue curves incorporate mis-centering by convolving the convergence profile with a beam-dependent Rayleigh distribution.}
\label{fig:cmbBeam}
\end{figure}

Our model for the CMB lensing convergence signal is an NFW profile with mass $M_{500}$, virial radius $R_{500}$ and concentration $c_{500}$ :

\begin{equation}                                                                                         
\kappa(\theta) = {3 \over 4\pi c^2} {H_0^2 \chi_L}{\chi_{LS} \over {\chi_S}} (1+z_L) {M_{500} c_{500}^2  \over \rho^c_0 R_{500}^2 } {g(\theta /\theta_s) \over f({c_{500}})}               
\end{equation}    
where $\theta$ is angle on the sky, $c$ is the speed of light, $H_0$ is the Hubble constant, $z_L$ is the redshift of the cluster, $\theta_s$ is the scale radius $\theta_s=R_{500}/c_{500}$, $\rho_0^c$ is the critical density of the Universe today, $\chi_S$ is the comoving distance to the CMB, $\chi_L$ is the comoving distance to the lens cluster, $\chi_{LS}$ is the comoving distance between the CMB and cluster,
\begin{equation}                                                                                         
f(c) = \ln (1+c) - {c \over 1+c}\,,                                                                      
\end{equation}    
and
\begin{equation}
g(x) =
    \begin{cases}
      { 1 \over x^2-1} \left[ 1- {2 \over \sqrt{x^2-1}} {\rm atan} \sqrt{ {x-1 \over x+1}}\right],  \quad (x>1)\\
      { 1 \over x^2-1} \left[ 1- {2 \over \sqrt{1-x^2}} {\rm atanh} \sqrt{ {1-x \over1+x}}\right], \quad \
(x<1)\\
      { 1 \over 3}\,,  \quad (x=1)\,.
    \end{cases}       
\end{equation}

The relative error on the lensing mass of a given cluster is

\begin{equation}
{\sigma(M) \over M} = {\sigma(\kappa_{5\theta_{500}}) \over \kappa_{5\theta_{500}}}
\end{equation}
where $\kappa_{5\theta_{500}}$ is the integrated convergence within a disk of radius $5\theta_{500}$. The matched filter variance in the same region is given by

\begin{equation}
{\sigma^{-2}(\kappa_{5\theta_{500}}) = \int \dd^2\boldsymbol{L} { U(\boldsymbol{L})U^{\star}(\boldsymbol{L}) \over C^{\kappa\kappa}_{L}+N^{\kappa\kappa}_{L}}}
\end{equation}
where

\begin{equation}
U(\theta) = {\kappa(\theta) \over \kappa_{5\theta_{500}}}.
\end{equation}
Here $C^{\kappa\kappa}_{L}$ is the power spectrum of the convergence field (a line-of-sight integral over the cosmological matter power spectrum including non-linear corrections from Halofit) which captures fluctuations in the lensing field that are not related to the NFW cluster, and $N^{\kappa\kappa}_{L}$ is the lensing reconstruction noise per mode described earlier. We note that at low instrument noise levels, the quadratic estimator is less optimal than maximum likelihood techniques \cite{Raghu} causing an underestimate of sensitivity. However, in the small lens limit, the approximation in Eq 26 that the noise modes are uncorrelated also breaks down resulting in an overestimate of sensitivity \cite{Horowitz}. Since most clusters in the cosmological sample are not in this regime of high S/N and since the information on halo masses does not purely come from the small lens limit, we ignore these complications and leave a more complete treatment for later work.

We can now compare the performance of CMB lensing and optical lensing. Using the formalism described above, in Figure~\ref{fig:cmbVoptical}, we compare the signal-to-noise-ratio per cluster $\sigma(M) \over M$ for clusters of various masses for a 1 arcminute FWHM beam experiment that utilizes both temperature and polarization for CMB lensing (and includes noise from temperature foregrounds). Since shape noise increases with redshift as fewer source galaxies become available, optical weak lensing starts becoming less statistically informative for clusters at redshifts greater than around $z=1.2$ depending on the mass of the cluster.

When stacking on clusters where we have no optical follow-up data (for the highest-redshift clusters), assuming that the tSZ centroid is the center of the cluster can result in a smearing of the signal. To model this, we convolve the convergence profile with a Rayleigh distribution:

\begin{equation}
P(\theta) = {\theta \over \sigma_m^2}\mathrm{exp}(-{1 \over 2}{\theta^2 \over \sigma_m^2})
\end{equation}
where $\sigma_m$ is taken to be half the FWHM of the beam.

While the resolution of the CMB instrument affects the number of clusters detected through the tSZ effect, it also affects the mass sensitivity since a higher resolution experiment images smaller scales in the CMB temperature and polarization field that contribute to the lensing signal. In Figure~\ref{fig:cmbBeam}, we show the dependence of the mass sensitivity on beam FWHM for five scenarios. The most optimistic assumes that both temperature and polarization data are used and that there are no sources of noise from foregrounds (discussed in Section IIA). Foregrounds in temperature degrade the mass sensitivity by up to 80\%. If one assumes there is no foreground contamination in polarization, utilizing only the EE and EB polarization based estimators results in further degradation of mass sensitivity by around 20\%. While galactic foregrounds in polarization uncorrelated with the positions of galaxy clusters are highly uncertain at small-scales, polarized emission from clusters is expected to be well below the 1$\mu K$ level \cite{LouisPol}.   We also show in Figure~\ref{fig:cmbBeam} how mis-centering enhances the degradation of sensitivity as a function of beam FWHM. Our baseline forecasts assume both temperature and polarization data with foregrounds in temperature, and mis-centering only for clusters with $z>2$.

\begin{figure}[t]
\includegraphics[width=0.95\columnwidth]{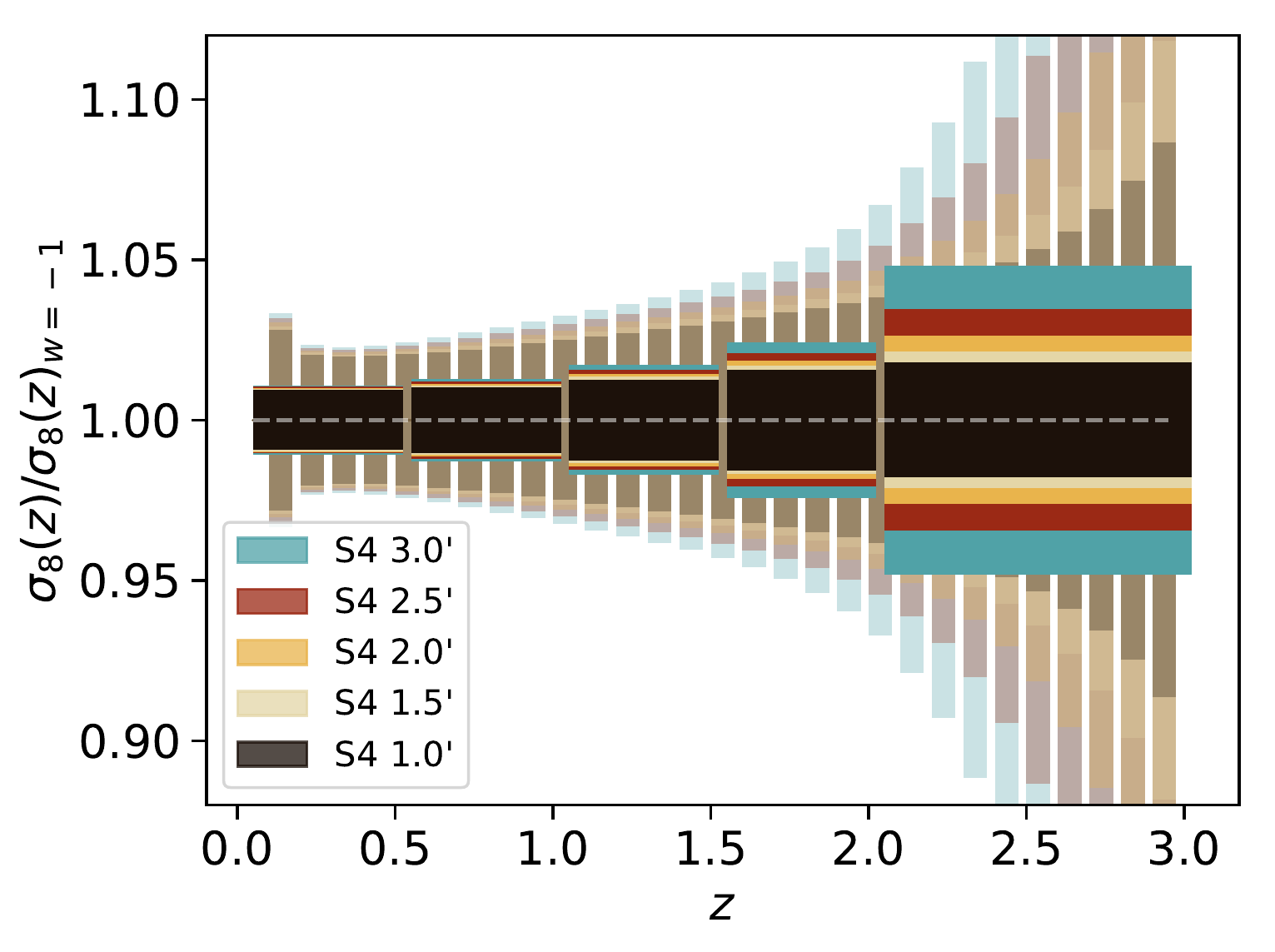}
\caption{The uncertainty on $\sigma_8(z)$ as a function of redshift for tSZ clusters from a CMB Stage-4 experiment calibrated internally using CMB halo lensing (including temperature and polarization data). The dark blocks have redshift bin edges of $[0.,0.5,1.0,1.5,2.0,3.0]$ chosen to roughly produce the same relative constraint in each bin, while the light blocks illustrate the $\sigma_8(z)$ constraints for bins of width $\Delta z = 0.1$. The different colors correspond to different beam FWHMs at 150 GHz. With a 1 arcminute beam at 150 GHz it is possible to constrain $\sigma_8(z)$ at $\sim 1$\% in the redshift bins chosen above.}
\label{fig:s8}
\end{figure}

\begin{table}
 \label{tab:steps}
 \begin{center}
  \caption[Parameter Values]{The fiducial and step values used in calculating derivatives for the Fisher matrix. The step sizes have been checked for numerical stability. We adopt a prior on the optical depth of 0.01 and restrict the overall weak lensing mass calibration to 1\% by allowing it to float and adopting a prior of 1\%. }
  \begin{tabular}{lccc}
   \hline \hline
Parameter         & Fiducial & Step & Prior \\ \hline
$\Omega_c h^2$    & 0.1194         & 0.0030 & \\
$\Omega_b h^ 2$   & 0.022        & 0.0008 & \\
$H_0$ & 67.0        & 0.5 & \\
$10^9A_s$             & 2.2    & 0.1 & \\
$n_s$             & 0.96         & 0.01 & \\
$\tau$            & 0.06          & 0.02 & 0.01 \\
\hline
$\Sigma m_\nu$ (meV)   & 60           & 20 & \\
$w_0$             & -1             & 0.05 & \\
$w_a$             & 0              & 0.1 & \\
\hline
$b$             & 0.8              & 0.02 & \\
$\alpha_y$             & 1.79              & 0.04 & \\
$\sigma_{\mathrm{log}Y}$             & 0.127              & 0.02 & \\
$\gamma_Y$             & 0              & 0.02 & \\
$\beta_y$             & 0              & 0.02 & \\
$\gamma_\sigma$             & 0              & 0.02 & \\
$\beta_{\sigma}$             & 0              & 0.02 & \\
\hline
$b_{\mathrm{WL}}$ & 1 & 0.1 & 0.01\\
$\sigma_m$ & 0.75 & 0.2 & beam/2 \\

\hline
  \end{tabular}
 \end{center}
  \begin{quote}
    \noindent 

    \end{quote}
\end{table}

\section{Cosmological Constraints}
\label{sec:res}
By finding galaxy clusters and calibrating their masses, we are constraining the halo abundance $n(M,z)$, a function that is sensitive to the amplitude of matter fluctuations $\sigma_8$ and the total matter density $\Omega_m$. The halo abundance is related to the matter power spectrum (and consequently the growth factor). Physics that affects the matter power spectrum or its growth can lead to differing predictions on the abundance of halos as a function of mass and redshift.

\begin{figure*}[t]
\includegraphics[width=\textwidth]{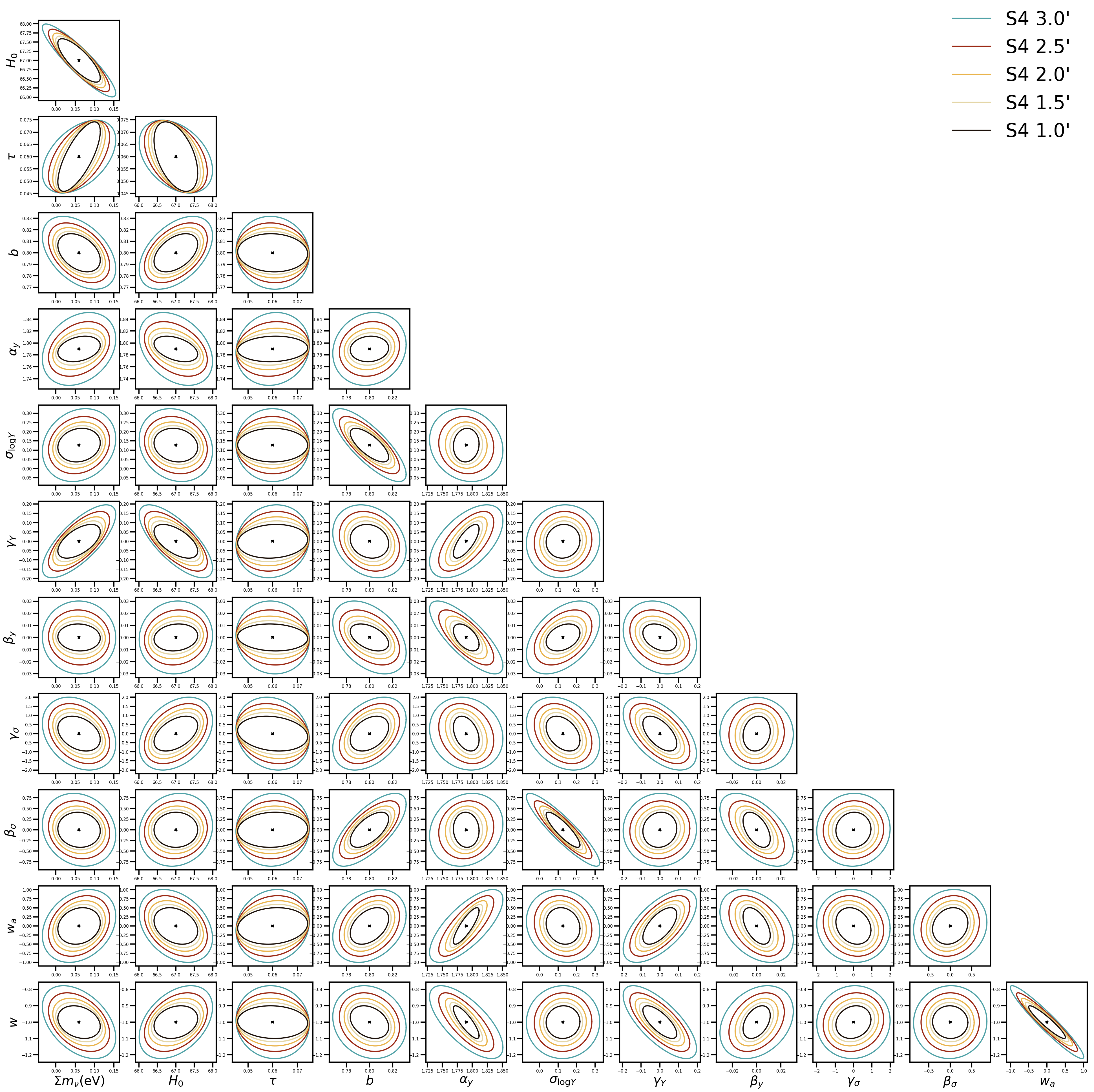}
\caption{Resolution dependence of CMB Stage-4 parameter constraints for internal CMB lensing calibration with temperature and polarization data for a $\Lambda CDM+m_{\nu}+w_0+w_a$ cosmology. The contours correspond to 68 \%C.L. levels. For clarity, not all parameters varied are shown -- $\{\Omega_{c}h^2,\Omega_{b}h^2, A_s,n_s\}$ are excluded.}
\label{fig:allparams}
\end{figure*}

\begin{figure*}[t]
\includegraphics[width=0.48\textwidth]{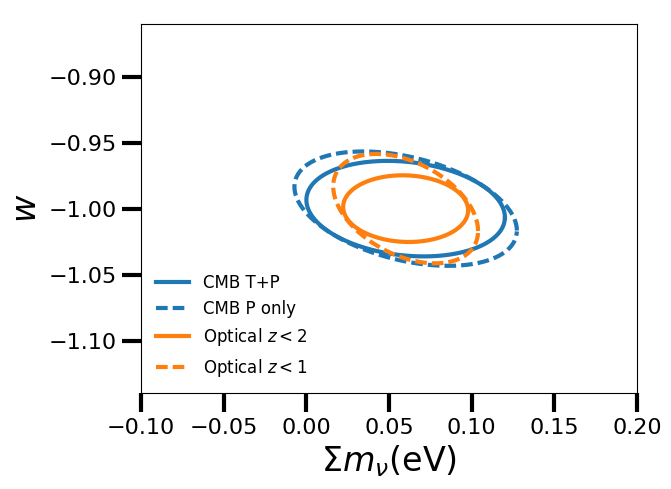}
\includegraphics[width=0.48\textwidth]{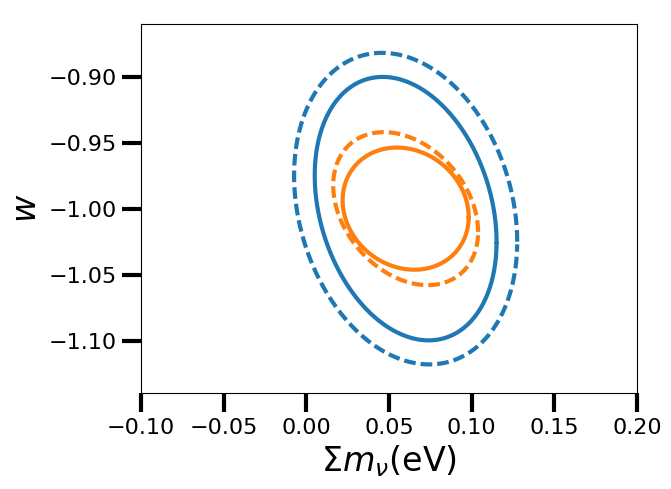}
\caption{Constraint in the $m_{\nu}-w_0$ plane for a CMB Stage-4 telescope with $1'$ beam FWHM at 150 GHz shown as 68 \%C.L. levels.  {\it Left:} For a $\Lambda CDM+m_{\nu}+w_0$ cosmology with $w_a=0$ held fixed. {\it Right:} For a $\Lambda CDM+m_{\nu}+w_0+w_a$ cosmology. The blue contours are for mass calibration using CMB lensing (temperature and polarization in solid and polarization only in dashed). The orange contours use mass calibration from an LSST-like optical survey. The blue dashed contour uses CMB polarization only, and the dashed orange contour only uses $z<1$ source galaxies.}
\label{fig:owlcmb}
\end{figure*}

\begin{figure*}[t]
\includegraphics[width=0.325\textwidth]{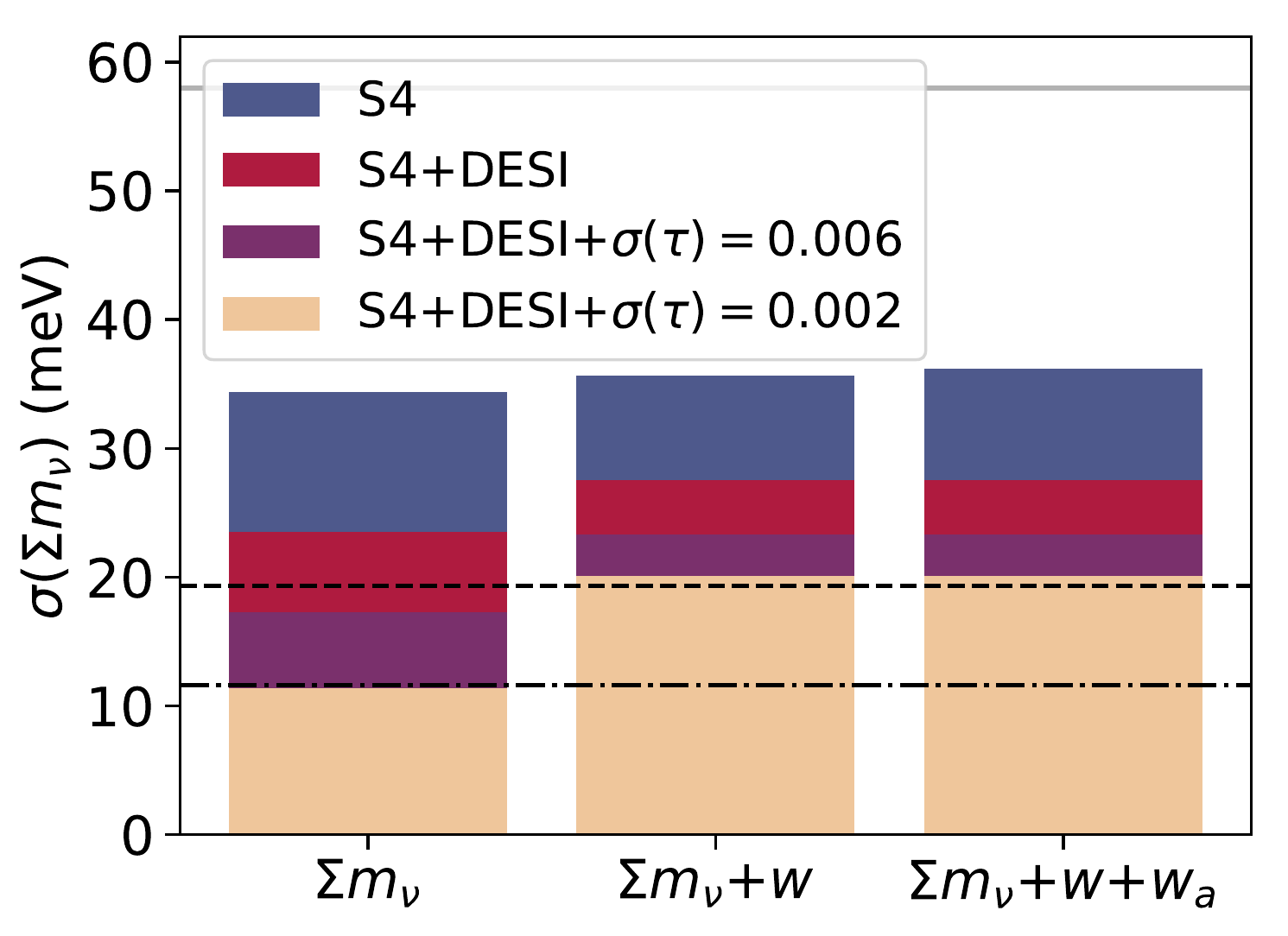}
\includegraphics[width=0.325\textwidth]{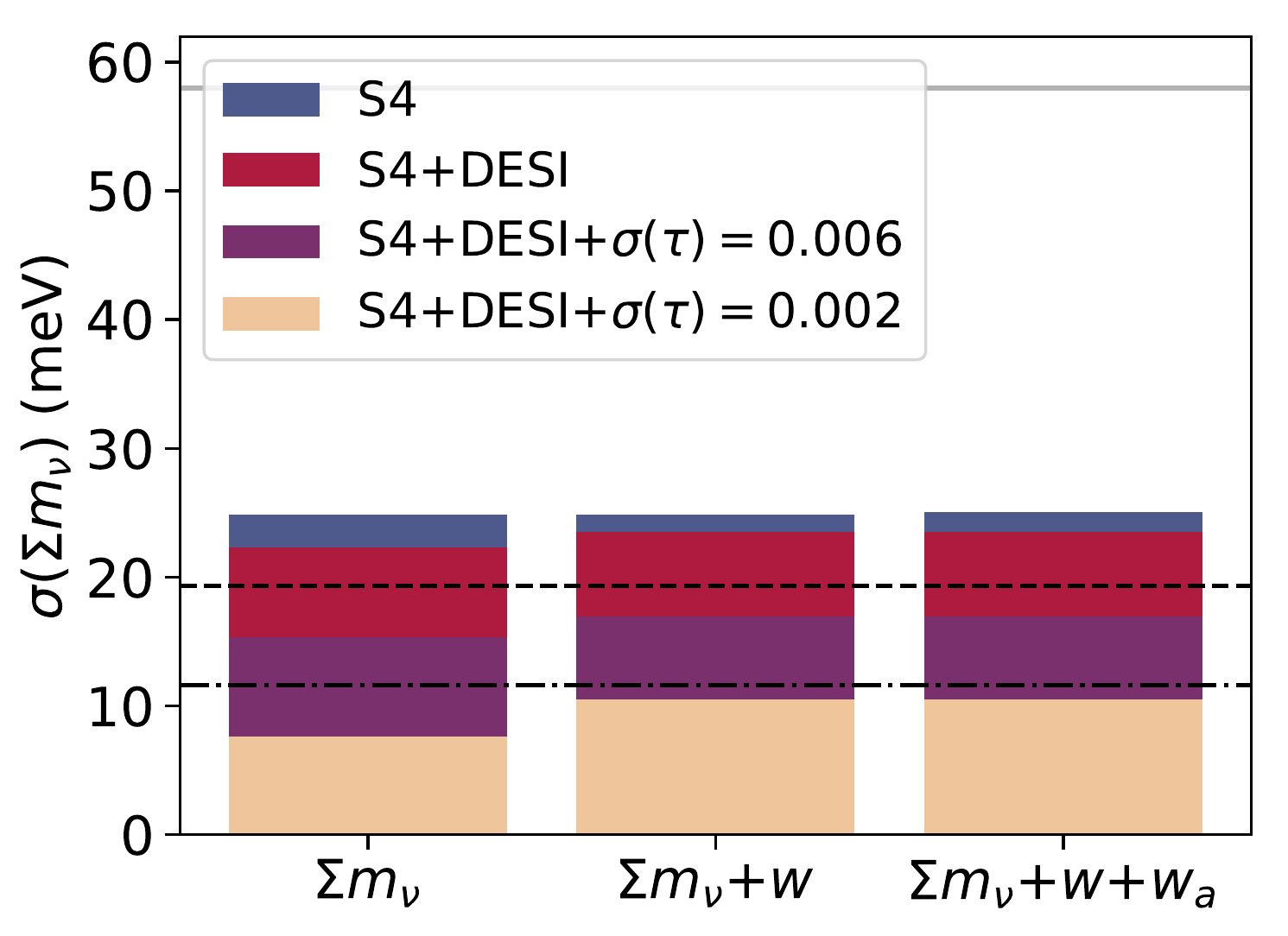}
\includegraphics[width=0.325\textwidth]{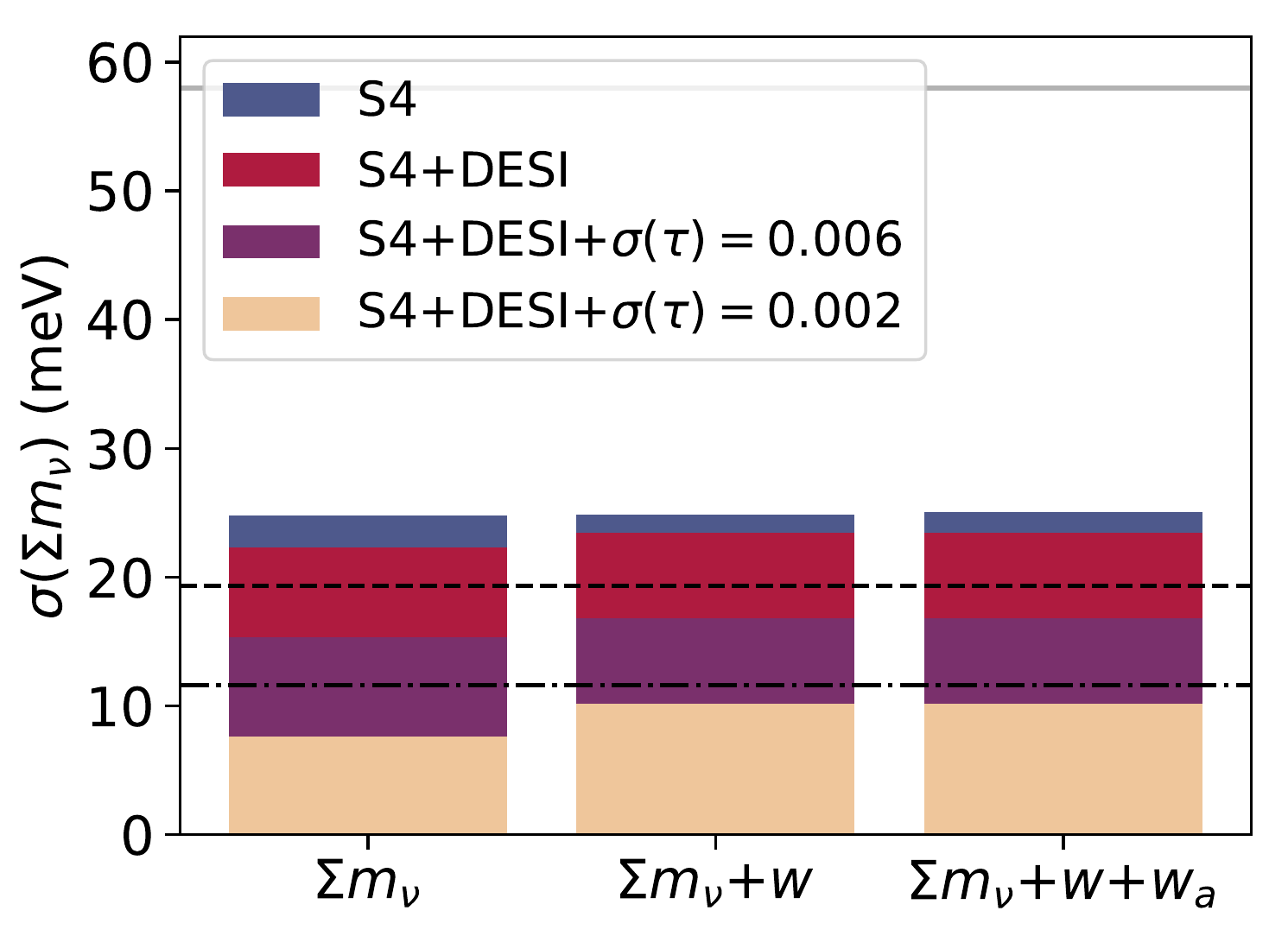}
\caption{The $1-\sigma$ uncertainty on neutrino mass obtained when marginalizing over $\Lambda CDM$, $\Lambda CDM+w_0$ and  $\Lambda CDM+w_0+w_a$, from tSZ clusters detected using a CMB Stage-4 telescope with $1'$ beam FWHM at 150 GHz . Unlike in the case where only the CMB lensing auto spectrum is used \citep[see][]{Allison2015}, the constraints do not degrade significantly when freeing up dark energy equation of state parameters owing to the redshift resolution of growth of structure with tSZ clusters. {\it Left:} Constraints when the mass calibration is from internal CMB lensing reconstruction with temperature and polarization data. {\it Middle:} Constraints when the mass calibration is from an LSST-like optical survey using clusters up to $z=2$. {\it Right:} Constraints when the mass calibration is a combination of internal CMB and optical weak lensing. Note that ``DESI'' corresponds to adding BAO measurements from the DESI survey. We show how constraints improve when the prior on the optical depth $\tau$ is tightened from the fiducial width $0.01$ to the Planck Blue Book value \citep{PlanckBB} of $0.006$ and further to the cosmic-variance-limit value of $0.002$. The grey solid line shows the value of the minimal neutrino mass in the normal hierarchy of 58 meV, and the dashed and dot-dashed lines show levels required for a 3$\sigma$ and 5$\sigma$ detection respectively.}
\label{fig:mnubar}
\end{figure*}

We obtain predicted constraints on cosmological parameters by calculating Fisher matrices for each experimental configuration \citep[e.g.,][]{Fisher1935,Knox1995,Jung1996}. We proceed as in \cite{ShimonEtAl,LA2017}, modeling the observed number counts $N$ as Poisson distributed about the predicted mean number of clusters $\bar{N}$ (See Equation~\ref{eq:abund}) in each $(M,q,z)$ bin,

\begin{equation}
\mathrm{ln} \mathcal{L}(N|\bar{N})=\sum_{M_L,q,z} N\mathrm{ln}\bar{N}-\bar{N}-\mathrm{ln}(N!)
\end{equation}
and obtain 68\% C.L. constraints by expanding assuming the likelihood is Gaussian in the parameters. This requires us to calculate derivatives of the observed number counts as a function of the parameters varied. The Fisher matrix is

\begin{equation}
F_{ab} = \sum_{M_L,q,z} {\partial_a N \partial_b N \over N}
\end{equation}

where $a$ and $b$ indicate the cosmological, scaling relation and other nuisance parameters. Marginalized 68 \%C.L. constraints on parameter $a$ for instance can then be calculated as

\begin{equation}
\sigma(a) = \sqrt{F^{-1}_{aa}}
\end{equation}

We vary $N(M,z,q)$ with respect to the following set of parameters. When including primary CMB information, our $\Lambda CDM$ parameter set is $\theta_c = \{\Omega_{c}h^2,\Omega_{b}h^2, H_0, A_s,n_s,\tau\}$. When not including primary CMB, we exclude $\tau$ from the Fisher matrix. In addition, we always marginalize over a set of scaling relation parameters $\theta_s=\{b,\alpha_y,\sigma_{\mathrm{log}Y},\gamma_Y,\beta_y,\gamma_\sigma,\beta_{\sigma}\}$ described earlier in Section III. The other parameters and external data sets we consider are described below.

Before these derivatives are calculated, the original $(M,q,z)$ grids are rebinned such that all clusters with $z>2$ are in a single bin and such that the $M$-bins are coarse enough (37 bins between $13.5<\mathrm{log}_{10}M<15.7$) given the mass calibration errors. 

\subsection{External data sets}

The cosmological constraints considered here will benefit from primary CMB information that pins down the amplitude of primordial power. For this purpose, we include a CMB Stage-4 Fisher matrix for $\{\theta_c\}+\Mnu$ when varying neutrino mass and $\{\theta_c\}$ when not. For CMB Stage-4, Fisher information is only included for the temperature multipole range $300<\ell<3000$ and the polarization multipole range $100<\ell<5000$. In addition, we include a Planck Fisher matrix for temperature and polarization. We avoid including low-$\ell$ polarization from Planck and avoid double counting as follows. We include $2<\ell<30$ Planck temperature information for $f_\rmn{sky}=0.6$, $30<\ell<100$ Planck temperature and polarization in the overlapping sky of $f_\rmn{sky}=0.4$ and $30<\ell<2500$ Planck temperature and polarization in the non-overlapping sky of $f_\rmn{sky}=0.6-0.4=0.2$. We use unlensed spectra to effectively exclude $\Mnu$ information from primary CMB. In lieu of including low-$\ell$ polarization from Planck, we impose a flat prior of 0.01 on $\tau$ (unless otherwise specified) whenever including primary CMB data.

Although cluster counts measure $\Omega_m$, some improvement in cosmological constraints can be obtained by the addition of baryon acoustic oscillations (BAO) measurement through the information it provides on $H_0$. We consider a BAO experiment like DESI and calculate its Fisher matrix. While not critical for the science targets, we show how much the addition of DESI can improve our constraints. The Fisher formalism for CMB and BAO and the experimental configurations for Planck and DESI follow those used in \citep{Allison2015}.

\subsection{Additional nuisance parameters}

We explore the effect of imposing a 1\% floor on the systematic uncertainty in the mass calibration since in the case of optical weak lensing a combination of shear multiplicative bias, photo-z uncertainties and modeling uncertainties is expected to lead to an overall floor at that level. We impose this floor by re-scaling the lensing mass as 
\begin{equation}
M_{L} \rightarrow b_{WL}M_{L}
\end{equation}
marginalizing over $b_{WL}$ but with a prior of 0.01. We find that this has very little effect on parameter constraints since the 37 mass bins used in the analysis allow for some self-calibration \citep{SelfCal} through information in the shape of the mass function. We do not vary $b_{WL}$ in the main results in this work.

For CMB lensing, we marginalize over a mis-centering offset $\sigma_m$ discussed in Section\ref{sec:cmbhalo} for clusters with $z>2$ and impose a prior on it that is equal to half the beam FWHM.

\subsection{Amplitude of the matter power spectrum $\sigma_8(z)$}
 
In order to project the sensitivity to the amplitude of matter fluctuations as a function of redshift, we calculate the following derivatives,

\begin{eqnarray}
\frac{\partial N(M,z_i,q)}{\partial \sigma_8(z_i)}I &=& \frac{1}{h}\{N[(1+h/2)^2P(k,z_i)] \nonumber \\
&-&N[(1-h/2)^2P(k,z_i)]\} .
\end{eqnarray}

where $h$ is the step size for the derivative calculation which we take to be $h=0.05$ (with other values tested to confirm stability).  When calculating these derivatives, we fix $\{\theta_c,\theta_s\}$ to their fiducial values varying only the power in each redshift bin that is used in the calculation of halo abundances. These derivatives are then stitched into a Fisher matrix for parameters $\{\theta_c,\theta_s,\sigma_8(z_i)\}$ and a CMB Fisher matrix for $\{\theta_c\}$ is added to it. When reporting constraints on the overall amplitude of power $\sigma_8$, we involve a single derivative that varies power across all redshifts.

In Figure~\ref{fig:s8}, we show constraints on $\sigma_8$ in redshift bins for the various telescope resolutions in Table~\ref{tab:cmbexp}, where we assume internal CMB lensing calibration with T+P. While the amplitude of the tSZ effect is roughly constant as a function of redshift, a higher resolution experiment is able to find tSZ clusters at higher redshifts that subtend smaller angles on the sky. These clusters are also calibrated better with CMB lensing as the resolution improves as indicated in Figures~\ref{fig:cmbVoptical} and~\ref{fig:cmbBeam}. The increased resolution thus primarily improves $\sigma_8$ constraints at higher redshifts. This improvement begins to saturate below 1.5 arcminutes due to degeneracies with other parameters (see~\ref{sec:con}).

\subsection{The sum of neutrino masses $\Mnu$}

Measurements of neutrino oscillations indicate that neutrinos are massive, but the absolute mass scale (sum of the masses of three neutrino species) is not known. Solar and atmospheric neutrino measurements allow for either a normal or inverted hierarchy of the three species with the minimal possible mass scales in each case being 58 meV and 100 meV respectively. Any cosmological observable sensitive to $P(k,z)$ in principle offers information on the neutrino mass scale since massive neutrinos become non-relativistic around $z=300$ and affect the growth of structure. In particular, the power spectrum is suppressed on scales smaller than the neutrino free-streaming scale.

The CMB lensing power spectrum (which depends on the integrated line-of-sight matter power) is likely the cleanest cosmological probe of $\Mnu$ since for CMB Stage 4 lensing constraints will be driven by polarization data with potentially fewer astrophysical systematics. However, because of a degeneracy with $\Omega_m h^2$, CMB lensing will need to be combined with external data from baryon acoustic oscillation (BAO) surveys like DESI in order to approach sensitivities capable of providing evidence for a minimal neutrino mass. the tSZ cluster abundances however are also highly sensitive to the neutrino mass scale since the neutrino free streaming scale below which matter fluctuations are suppressed is larger than the typical virial radii of even the most massive galaxy clusters. A cosmology with a large neutrino mass scale will therefore have fewer clusters. Importantly, cluster abundances also measure $\Omega_m h^2$, and therefore provide better constraints than CMB lensing alone in the absence of external data.

We allow for a non-zero sum of neutrino masses through its effect on the growth of structure (captured in the matter power spectrum $P(k,z)$) and consequently on the number density of detected clusters.

\subsection{The dark energy equation of state $w(a)$}

We consider the dark energy equation of state parametrized as

\begin{equation}
w(a) = w + (1-a)w_a
\end{equation}
and forecast constraints either on $\{w\}$ (with $w_a=0$) or on the combinations $\{w,w_a\}$, $\{\Sigma m_{\nu},w\}$ and $\{\Sigma m_{\nu},w,w_a\}$.

Measurements of the CMB lensing power spectrum are sensitive to the integrated matter power spectrum and lack redshift resolution. Since the equation of state of dark energy affects the growth of structure at low redshifts ($z<2$) and massive neutrinos suppress power below the free-streaming much earlier on, CMB lensing power spectra measurements suffer a degeneracy between $\Mnu$ and $w$ \citep{Allison2015}. Counting clusters in redshift bins significantly alleviates this degeneracy. In particular, as can be seen in Figure \ref{fig:allparams} higher resolution telescopes that find more clusters at higher redshifts suffer less degeneracy between $\Mnu$ and $w$.

In Figure \ref{fig:owlcmb}, we look at constraints in either a $\Lambda$CDM+\{$\Mnu$,$w$\} cosmology (left panel) or a $\Lambda$CDM+\{$\Mnu$,$w$,$w_a$\} cosmology (right panel). In the first case, allowing only $z<1$ clusters when optical weak lensing is used as a mass calibrator (anticipating photometric redshift systematics for source galaxies $z>1$) we find that CMB lensing performs comparably or better regardless of whether we restrict ourselves to polarization data only. When $w_a$ is freed up, optical weak lensing performs better relative to CMB lensing. Figure 8 looks at the marginalized constraint on the sum of neutrino masses alone, in cosmologies with $w_0$ fixed, $w_0$ varied with $w_a$ fixed, and both $w_0$ and $w_a$ varied. We find that in contrast to the CMB lensing power spectrum, constraints on neutrino mass are not significantly degraded when the equation of state of dark energy is freed up. 

\begin{figure}[t]
\includegraphics[width=0.95\columnwidth]{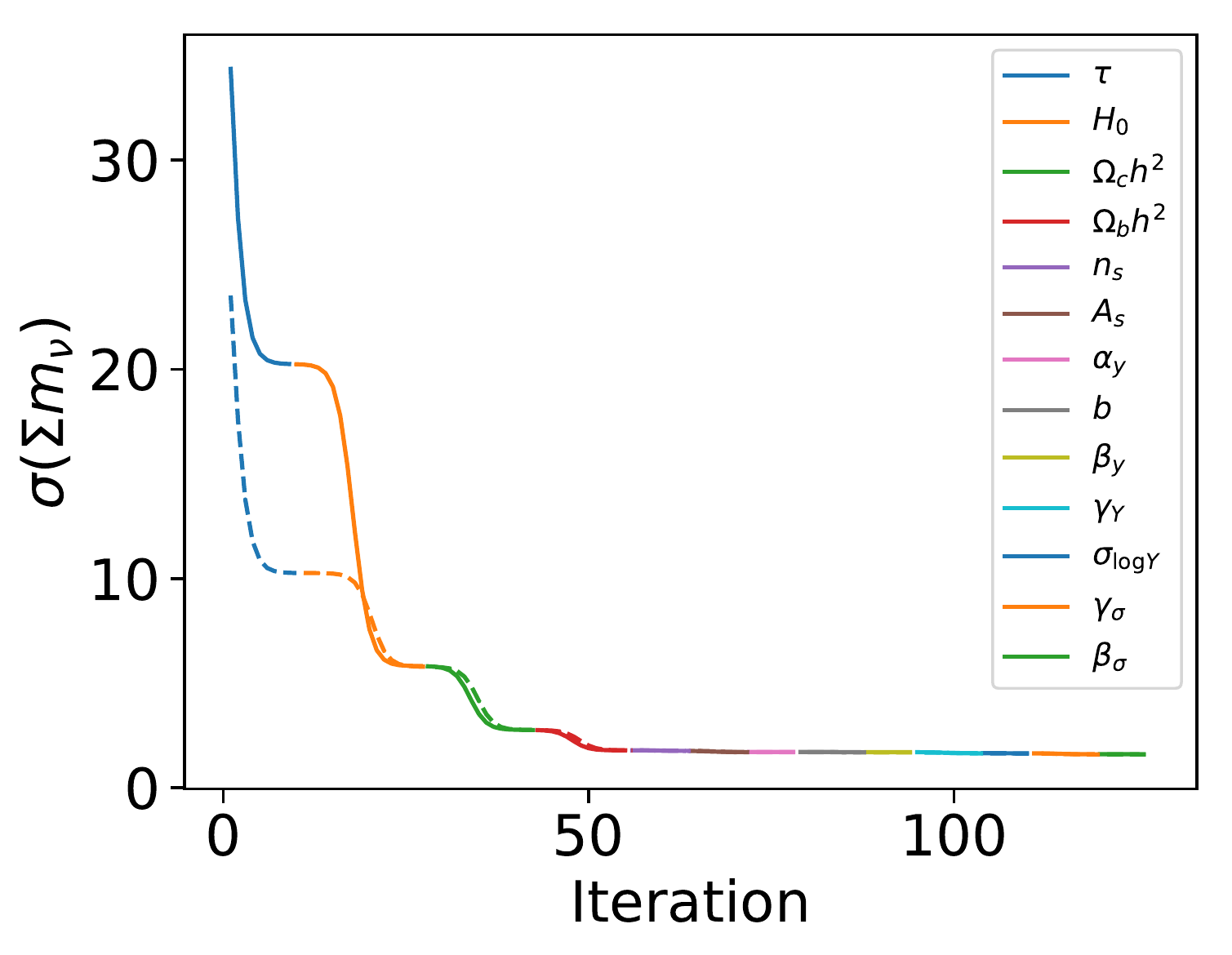}
\includegraphics[width=0.95\columnwidth]{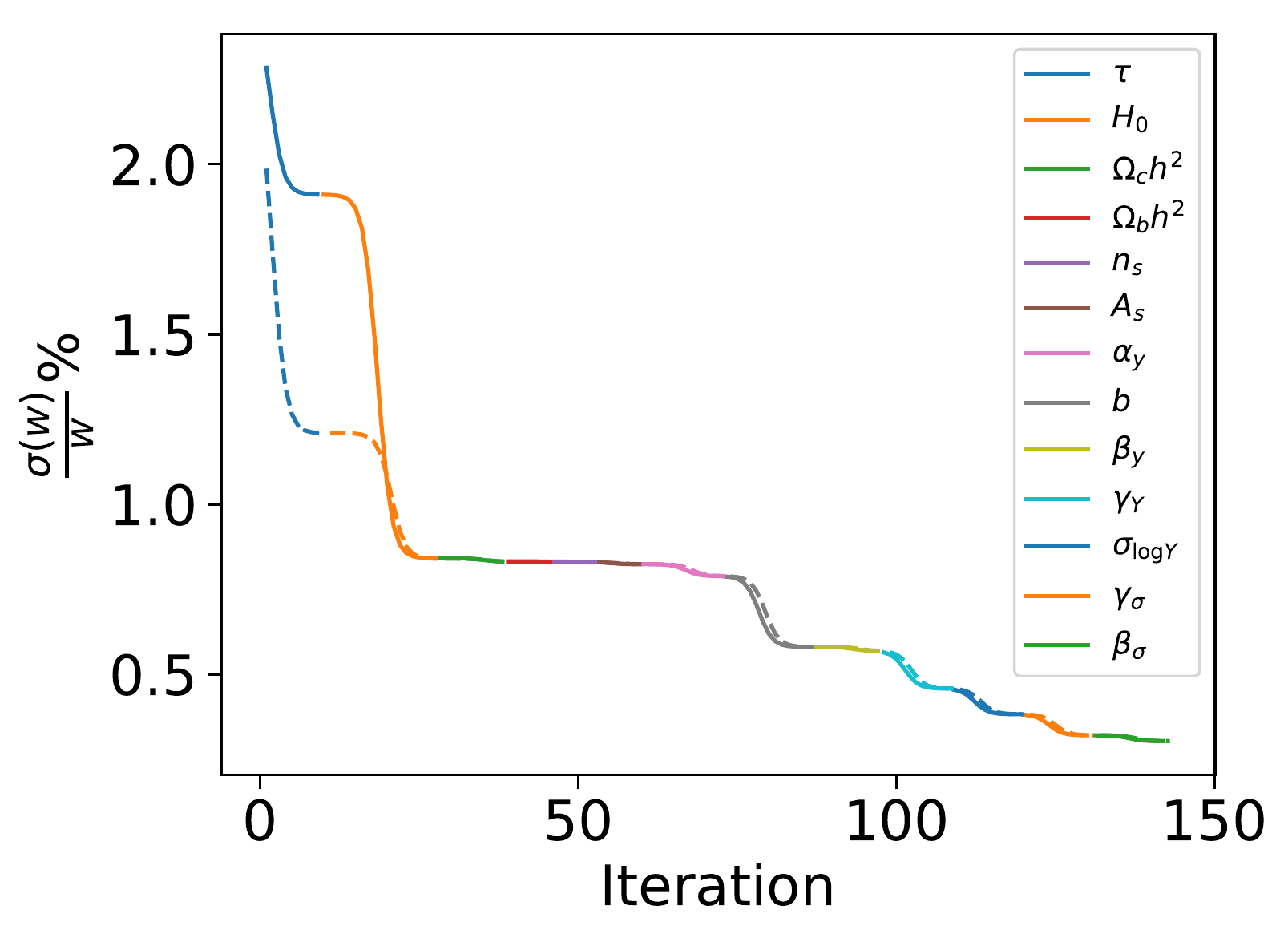}
\caption{The constraint on the dark energy equation of state and sum of neutrino masses as we increase prior information on each parameter for CMB Stage-4 (1 arcminute at 150 GHz) with internal CMB halo lensing (T+P) calibration. With each step in the forward-$x$ direction, we decrease the 68 \%C.L. width of the prior on the respective parameter on a logarithmic scale until it saturates the constraint (determined as the point when the change in constraint is less than 0.1\%). Once this saturation is seen, we retain the final value of the prior for that parameter and proceed to repeat the same procedure on the next parameter of interest (represented by a different color). This makes clear which parameter degeneracies are primarily limiting the constraint. The dashed curve includes DESI BAO information.}
\label{fig:prior}
\end{figure}

\section{Discussion and Conclusions}
\label{sec:con}
Future measurements of CMB secondary anisotropies from experiments like CMB Stage-4 will find a very large number of galaxy clusters, $N_c  \approx 100,000$, with the precise number depending on the resolution of the experiment. With such large statistics, many of the scaling relation parameters and the mass calibration self-calibrate. One would expect that the cosmological information would increase roughly with the square root of the number of clusters detected, especially since most of the clusters detected with higher resolution are at high redshifts. However, certain parameter degeneracies prevent there from being significant improvement in constraints beyond a resolution of around 1.5 arcminutes, particularly for the sum of neutrino masses.

Our fiducial forecasts assume knowledge of the optical depth to a precision of 0.01 (realized by the most recent Planck analysis of low-ell polarization \cite{PlanckTau}). Since a measurement of low-redshift amplitude of structure requires the amplitude of scalar fluctuations as a reference for untangling the effects of neutrino mass or dark energy, a degeneracy with the optical depth is introduced when using primary CMB data. This degeneracy is particularly limiting for $\Mnu$, and prevents increased cluster counts from higher resolution CMB measurements from commensurately improving constraints. In Figure~\ref{fig:mnubar}, we show the improvement obtained if the constraint on optical depth were 0.006 (the Planck Blue Book value) and the cosmic-variance-limited value of 0.002. Such improvements would allow the minimal neutrino mass in the normal hierarchy to be detected at the 5-sigma level or greater using CMB Stage-4 clusters + primary CMB data along, even when freeing up dark energy equation of state parameters, allowing for critical cross-checks of measurements made through other probes such as the CMB lensing power spectrum. The optical depth could be measured to the cosmic variance limit (CVL) by a future space-based CMB experiment or by a ground-based CMB experiment if significant advances are made in controlling low-$\ell$ polarization systematics. It could also be potentially improved beyond CVL with a 21 cm experiment \citep{Tau21}.  

Our knowledge of other parameters such as the Hubble constant is also a limiting factor in these constraints. We explore the contributions of all parameter degeneracies in Figure~\ref{fig:prior} by starting with our fiducial constraint and cumulatively improving the prior knowledge on every other parameter along the x-axis, switching to the next parameter once the previous parameter saturates the constraint. Perfect knowledge of $\tau$ and $H_0$ significantly improves constraints for $\Mnu$ and $w$ as expected.

Previous forecasts and analyses of cosmological constraints from CMB Stage-4 SZ clusters have come to same conclusions as this work with regards to the potential cosmological utility of SZ selected clusters. The constraints on $\Mnu$ shown in this work, \citet{LA2017}, and \citet{Melin2017} are competitive with other large-scale structure probes, for example CMB lensing \citep[][and references therein]{cmbs4}. These constraints have the advantage that they do not require an additional constraint on $\Omega_m h^2$. This work goes beyond previous work \citep[e.g.,][]{LA2017,Melin2017} in a few ways. First, we include the atmosphere in the noise modeling, which affects the number of clusters one expects to find (see Figure~\ref{fig:atm}). Second, we include additional degrees of freedom in the Y--M scaling relation modeling, such as mass and redshift dependent intrinsic scatter, and we explore a systematic uncertainty floor of 1\% on the lensing mass calibration. Thirdly, we include the unresolved tSZ source background, which is an irreducible source of noise for the match filter, since it has the same frequency dependence as the SZ clusters we are looking for. We consider the optical weak-lensing mass calibration from a conservative LSST-like survey over the entire CMB Stage-4 survey area, which will already be completed before CMB Stage-4 finishes its surveying. Finally, our cosmology constraints also extend to a model-independent forecast on the amplitude of matter fluctuations $\sigma_8(z)$ which demonstrates the effectiveness of larger telescopes in distinguishing between differing predictions of the growth of structure at high redshifts.

In this work we have been very conservative in how we implement the observable--mass relation, scatter, and the mass calibrations, however, if tSZ cluster counts are going to drive the design of CMB Stage-4 then moving beyond Fisher forecasts is imperative. Important steps forward include the incorporation of correlations between the weak-lensing and tSZ mass proxies. As expected these correlations are present \citep{Stanek2010,White2010,Shir2016}, and to properly quantify these correlation requires hydrodynamic cosmological simulations. Additionally, correlated foregrounds, in particular the tSZ-CIB correlation will need to be properly addressed with future analyses of cosmological simulations, similar to the analyses in \citet{Melin2017}. A realistic treatment of internal CMB halo lensing requires the full utilization of mutli-frequency information, explicitly projecting out the foregrounds that cause the most bias, a full accounting of the impact of kinetic SZ that cannot be removed through component separation and a comparison with maximum likelihood techniques that takes into account the presence of foregrounds. Finally, we do not consider any uncertainties on the halo mass function that could be as high as 10\% when considering the baryonic effects \citep[e.g.,][]{Cui2012,Cui2014,Bocq2016}.  Fortunately, these baryonic effects can be separated from the impact of $\Mnu$ on the halo mass function \citep{Mum2017}. Regardless of these proposed forecasts advancements that will require more sophisticated treatments like simulations, the trends with aperture size and the overall constraints, particularly on $\Mnu$, will not be affected substantially, since they are limited by external parameters like $\tau$.

\acknowledgments

We thank Steve Allen, David Alonso, Jim Bartlett, Brad Benson, Tom Crawford, Jo Dunkley, Colin Hill, Arthur Kosowsky, Thibaut Louis, Neelima Sehgal and David Spergel for their productive discussions and comments on this work. NB acknowledges the support from the Lyman Spitzer Jr. Fellowship. The Flatiron Institute is supported by the Simons Foundation. HM is supported by the Jet Propulsion Laboratory, California Institute of Technology, under a contract with the National Aeronautics and Space Administration.

\bibliographystyle{apj}
\bibliography{msm}

\end{document}